\documentclass[reprint,nofootinbib,superscriptaddress,tightenlines]{revtex4-1}

\usepackage[utf8]{inputenc}
\usepackage{amsmath}
\usepackage{amsfonts}
\usepackage{amssymb}
\usepackage{epsfig}
\usepackage{float}
\usepackage{graphicx}
\usepackage[normalem]{ulem}
\usepackage{color}
\usepackage{comment}
\usepackage[dvipsnames]{xcolor}
\usepackage{hyperref}
\hypersetup{
	colorlinks=true,
	linkcolor={red!90!black},
	citecolor={blue!90!black},
    urlcolor={blue!80!black}
}
\newcommand{\Teff}{T_\text{eff}}
\newcommand{\Gij}{G_{\mathbf{r}_{ij}}^{\psi_i}}

\newcommand{\Tmct}{T_\text{MCT}}
\newcommand{\cp}{\chi_4^\text{peak}}

\begin{document}

\title{An elastoplastic model approach for the relaxation dynamics of active glasses}

\author{Tanmoy Ghosh}
\email{gtanmoy@tifrh.res.in}
\affiliation{Tata Institute of Fundamental Research, Hyderabad - 500046, India}

\author{Peter Sollich}
\email{peter.sollich@uni-goettingen.de}
\affiliation{Institute for Theoretical Physics, University of G{\"{o}}ttingen, Friedrich-Hund-Platz 1, 37077 G{\"{o}}ttingen, Germany}
\author{Saroj Kumar Nandi}
\email{saroj@tifrh.res.in}
\affiliation{Tata Institute of Fundamental Research, Hyderabad - 500046, India}

\begin{abstract}
	
How activity affects the glassy dynamics is crucial for several biological processes. Furthermore, active glasses offer fascinating phenomenologies, extend the scope of equilibrium glasses, and can provide novel insights into the original problem. We introduce a family of novel approaches to investigating the relaxation dynamics of active glasses via an active elastoplastic model (EPM). These approaches describe the relaxation dynamics via local plastic yielding and can provide improved insights as we can study various aspects of the system separately. Activity enters the model via three crucial features: activity-mediated plastic yielding, activated barrier crossing, and persistent rotational dynamics of the yielding direction. We first consider a minimal active EPM that adds the effect of active yielding to a thermal EPM. We show that this active EPM captures the known results of active glasses within a reasonable parameter space. The results also agree well with the analytical results for active glasses when activity is small. The minimal model breaks down at very low temperatures where other effects become important. Looking at the broader model class, we demonstrate that whereas active yielding primarily dominates the relaxation dynamics, the persistence of yielding direction governs the dynamic heterogeneity in active glasses.
\end{abstract}
\maketitle

\section{Introduction}

Glassy properties of several biological systems, such as intracellular phase-separated biomolecular condensates \cite{jawerth2018,jawerth2020,alshareedah2021}, the cytoplasm \cite{fabry2001,deng2006,bursac2005,parry2014}, cell monolayers and tissues \cite{angelini2011,park2015,atia2018,malinverno2017}, and collections of organisms \cite{takatori2020,lama2024} are fundamental for various biological processes \cite{berthier2019,janssen2019,activematterreview}. Examples include wound healing \cite{poujade2007,das2015}, embryogenesis \cite{friedl2009,tambe2011,mongera2018}, cancer progression \cite{wirtz2011,park2016,mitchel2020,
streitberger2020} etc. Glassy dynamics refers, in the simplest notion, to a dramatic slowdown without much discernible change in static properties \cite{berthier2011,debenedetti2001}. It is accompanied by several dynamical features, such as complex stretched-exponential relaxation \cite{angelini2011,park2015}, sub-diffusive mean-square displacement \cite{bursac2005,deng2006}, non-Gaussian particle displacement distributions \cite{malinverno2017,trepat2009} and dynamical heterogeneity \cite{zhou2009,angelini2011}. Biological systems add a novel dimension to the original problem of glassiness, with various new control parameters and novel phenomenology. One crucial feature of these systems is activity, where the individual entities can self-propel with force $f_0$ and persistence time $\tau_p$ for their motion \cite{marchetti2013, ramaswamy2010, bechinger2016,vicsek2012}. Experiments, simulations, and theoretical work have shown that the glassy dynamics in these systems, referred to as active glasses, show non-trivial behavior as $f_0$ and $\tau_p$ vary \cite{angelini2011,garcia2015,nishizawa2017,mandal2016,flenner2016,berthier2013,szamel2016,nandi2017,nandi2018,paul2023,keta2023,sadhukhan2024,mandal2020PRL,mandal2021,keta2022,zheng2024,kolya2024}. Active glasses have enriched and broadened the field; we must now think about glassy dynamics from a more global perspective, including the aspects of biological systems that make the problem even more challenging but also more fascinating to study.

Despite much conceptual and technical progress in the recent past, the precise mechanism behind glassy dynamics remains an open question \cite{berthier2011,langer2014,biroli2022}. We know that relaxation in these systems results from local plastic rearrangements at particule level \cite{lemaitre2009,lemaitre2014,martens2012,ferrero2014,lin2014,nicolas2018}. Since thermal effects lead to more complex barrier-crossing scenarios, one can aim to first understand these events at zero temperature, inducing them athermally by external means such as shear \cite{martens2012,nicolas2018,lin2014}. Since the plastic events are local, their time scale is much shorter than the relaxation time of the system. Therefore, we can treat the system as an elastic medium on this time scale \cite{dyre2006,chattoraj2013,dyre2024,
schroder2020}; this implies that plastic events will affect the rest of the system via long-range elastic interactions \cite{eshelby1957,martens2012,nicolas2018}. Even at this level of simplification, the problem remains complex and computationally costly to investigate with particle-scale resolution. One can therefore further simplify by coarse-graining the system into mesoscopic regions $i$ and tracking the evolution of the local (shear) stresses $\sigma_i$, thus obtaining an elastoplastic model (EPM) \cite{bulatov1994,nicolas2018,ozawa2018,rossi2023}. In an EPM each plastic events leads to a local stress drop, and the elastic medium redistributes the excess stress to other sites. Yielding is almost always taken as incompressible, causing a local plastic strain with (in 2d) one extensional and compressional direction.

EPMs have until recently been primarily applied to athermal yielding \cite{martens2012, picard2004,lin2014,rossi2023}, where shear sets the direction of local deterministic yielding when $\sigma_i>\sigma_c$, where $\sigma_c$ is a critical value. In recent work, Ozawa and Biroli \cite{ozawa2023, tahaei2023} have removed these restrictions and developed an EPM for thermal yielding to study the relaxation dynamics in disordered systems: each plastic yield has a randomly oriented extension and compressional direction, and it is an activated event, i.e.\ a site can yield even when $\sigma_i<\sigma_c$, with a probability that depends on the distance from the yield value \cite{popovic2021, ferrero2021, ferrero2014}. Within this model, glassy relaxation occurs via dynamical facilitation, where yielding at one site facilitates relaxation at another. Furthermore, Ozawa and Biroli demonstrated that dynamical heterogeneity (DH) results from this facilitation process \cite{ozawa2023}. In this work, we extend this model to active glasses. Since activity introduces novel control parameters, our active EPM can provide crucial insights into the non-trivial features of these systems. We identify three distinct processes that can be affected by activity. We demonstrate that even simple members of the resulting model class reproduce the relaxation dynamics observed in active glasses. Interestingly, in contrast to equilibrium systems, the long-range facilitation alone is not enough to produce DH in active glasses; the rotational diffusivity of the yielding direction is also crucial.

\section{Results}

\begin{figure}
\includegraphics[width=8.4cm]{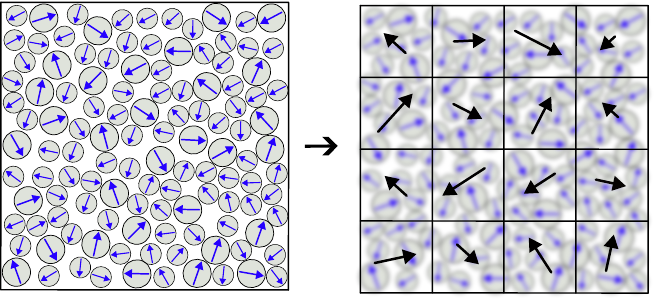}
\caption{Schematic representation of the active EPM. We divide the system into small meoscopic regions indicated by the boxes and coarse-grain to a representation where each box has a stress and a yielding direction, represented respectively by the magnitude and direction of the arrows.}
\end{figure}

\subsection{The active EPM}
We first develop the active EPM to study the relaxation dynamics in active glasses. We coarse-grain the system and divide it into $N=L^d$ mesoscopic blocks, where $L$ is the length of the system in units of block size, and $d$ is the spatial dimension. In this work, we have studied a two-dimensional model, i.e.\ $d=2$. We associate each block with a local shear stress $\sigma_i(t)$ at time $t$, where $i\in \{1,N\}$. We have studied the scalar EPM, where we keep track of the magnitude of the stress alone. We consider activity in the form of active Brownian particles (ABP) \cite{activematterreview}. Activity can have three main contributions to the relaxation process. 

First, active motion can facilitate yielding by generating active stresses. We incorporate this contribution via a fluctuating stress rate, $b_i$, at each site $i$ (equivalently an active shear rate times a local shear modulis). This means that, between discontinuous changes caused by plastic yielding, the stress at each site evolves according to 
\begin{equation}
\frac{\partial\sigma_i}{\partial t} = b_i.
\end{equation}
We assume that the active stress rates $b_i$ are spatially uncorrelated from one site to another and have Gaussian statistics with zero mean and covariance
\begin{equation}
\langle b_{i}(t) b_{j}(0)\rangle = \delta_{ij}\sigma_{0}^2  \exp[-t/\tau_p^b],
\end{equation}
where $\delta_{ij}$ is the Kronecker delta function, $\sigma_0$ is the strength of the stress rate and $\tau_p^b$ is its persistence time. The parameters $\sigma_0$ and $\tau_p^b$ are related  to the activity parameters of the individual particles, $f_0$ and $\tau_p$, but their scales will be different due to the coarse graining (see Sec. \ref{MCTcomparison}).

The second mechanism is yielding via barrier-crossing at an effective temperature, $\Teff=T+T_a$, where $T$ is the conventional equilibrium temperature and $T_a$ an additional active contribution \cite{benisaac2015,nandi2017,nandi2018}. Indeed, several recent studies \cite{woillez2019,caprini2019} have shown that active systems can go from one metastable minimum to another with a rate determined by $\Teff$, even when $T=0$. Within the active EPM, this leads to a non-zero probability of yielding via barrier crossing when $\sigma_i$ is less than $\sigma_c$: the yielding probability is $Y=\exp(-\Delta E_i/\Teff)$, where $\Delta E_i=(\sigma_c- |\sigma_i|)^a$. Consistent with MD simulations, we choose $a=3/2$ \cite{maloney2006}. On the other hand, a site undergoes deterministic yielding, that is, $Y=1$, when $|\sigma_i|>\sigma_c$. We have chosen $\sigma_c=1$ and used Monte Carlo method to simulate the yielding process \cite{ozawa2023}. 
After a successful plastic event at a site $i$, the stress decreases by $\delta\sigma_i = (z+|\sigma_i|-\sigma_c)\text{sgn}(\sigma_i)$, where sgn[\ldots] is the sign function and we draw $z$ from an exponential distribution, $\rho(z)=z_0^{-1} e^{-z/z_0}$ \cite{popovic2021,ozawa2023,lerbinger2022,barbot2018}. We set $z_0=1$ for simplicity. The stress on other sites, $j$, is updated as $\sigma_j\to\sigma_j+G_{\mathbf{r}_{ij}}^{\psi_i}\delta\sigma_i$, where $\Gij$ is the Eshelby propagator for stress \cite{eshelby1957}, $\mathbf{r}_{ij}$ is the spatial separation of the two sites, and the angle $\psi_i$ specifies the yielding direction, i.e.\ the extensional direction of the local plastic event.

The third and final effect of activity enters via the dynamics of the yielding direction $\psi_i$. As discussed in the introduction, we assume incompressible plastic events that have extensional and compressional axes. In the case of thermal yielding, $\psi_i$ is random since the direction of thermal yielding is random. However, activity can bring in a persistence time for the random variation of $\psi_i$; this angle then  performs rotational diffusion with a persistence time $\tau_p^{\psi}$,
\begin{equation}
\frac{\partial \psi_i}{\partial t} = \sqrt{\frac{2}{\tau_p^\psi}} \eta_i.
\end{equation}
Here $\eta_i(t)$ is a Gaussian white noise with zero mean and variance $\langle \eta_i(t) \eta_j(t') \rangle = \delta_{ij}\delta(t - t')$. The limit of small angular persistence time $\tau_p^\psi$ retrieves the thermal assumption of $\psi_i$ being random for each yield event.

Summarizing, the active EPM model includes activity via three distinct effects: yielding via an active stress rate, activated barrier crossing and the persistence of local yielding directions. The parameters $\sigma_0$, $\tau_p^b$, and $\tau_p^\psi$ come from the microscopic parameters of activity, $f_0$ and $\tau_p$. We expect the former set of parameters to be proportional to the latter. However, their detailed functional dependence on $f_0$ and $\tau_p$ is unclear.

\begin{figure*}
	\centering
	\includegraphics[width=12.6cm]{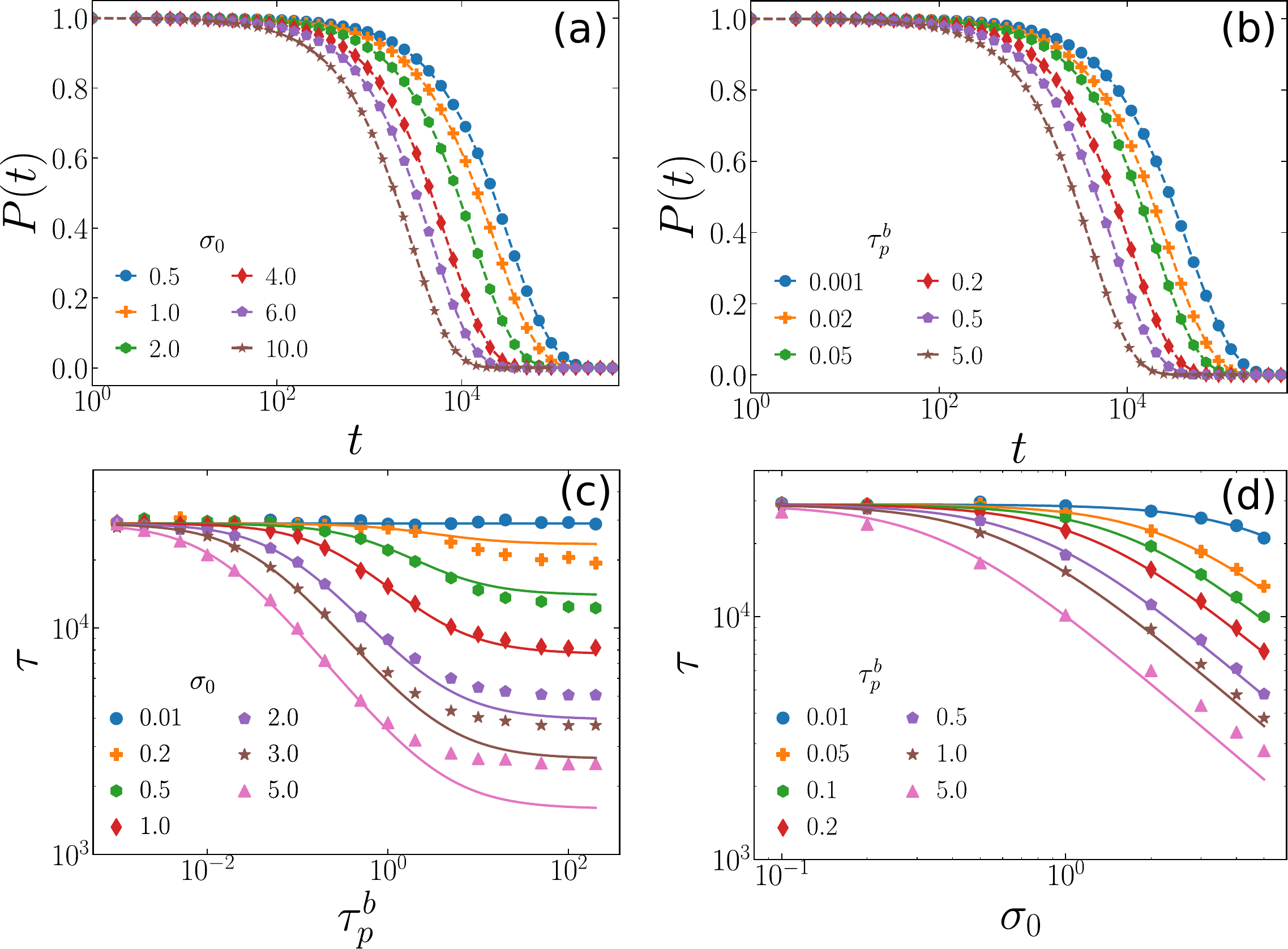}
	\caption{Relaxation dynamics for the thermal active EPM at $T=0.015$. (a) $P(t)$ decays faster with increasing $\sigma_0$ for fixed $\tau_p^b=1.0$. (b) $P(t)$ decays faster as $\tau_p^b$ increases for fixed $\sigma_0=5$. This behavior is consistent with ABP-type of activity as discussed in the text. (c) Relaxation time $\tau$ as a function of $\tau_p^b$ for different values of $\sigma_0$ and (d) 
		as a function of $\sigma_0$ for different $\tau_p^b$. Symbols are the simulation data, and dashed lines represent fits to the active MCT, Eq.~(\ref{mcttaufit}), with the constants $\mu = 0.5$, $A = 166.60$, $B =3.3$, and $C = 0.25$.}
	\label{fig-1}
\end{figure*}

We therefore first simplify to a minimal model that we call ``thermal active EPM''. This contains, beyond thermal yielding with random orientations ($\Teff=T$, $\tau_p^\psi\to 0$), only the active stress rate. We show that this already reproduces known simulation results with its two remaining activity parameters, $\sigma_0$ and $\tau_p^b$. We will subsequently examine the two additional effects and show that the effect of yielding via the active stress rate is indeed dominant for the relaxation. We will finally demonstrate that it is the rotational diffusivity of $\psi$ that controls DH in active systems.

\subsection{A minimal model for the relaxation dynamics in thermal active systems }
\label{minmod}

We consider a minimal active EPM and demonstrate that it already captures the main features of thermal active glasses. This minimal model adds to the thermal EPM of Ref. \cite{ozawa2023} the effect of the active stress rate $b_i$. The thermal EPM (at $T>0$) assumes random orientation of the yielding direction $\psi_i$ \cite{ozawa2023,tahaei2023}.
We expect that at very low $T$, 
there will be a competition between thermal activation and activity-induced persistence of $\psi_i$. 
But for the minimal model, we consider the regime of small activity where the persistence of $\psi_i$ can be treated as a negligible perturbation.
Similarly, we assume that the active contribution
$T_a$ to the effective temperature is much smaller than $T$ and modifies the rates of activated yielding in a negligible manner. Thus, we take $\Teff\simeq T$ for the thermal active system with $\tau_p^\psi=0$. This simplification 
facilitates comparison with the existing simulation and theoretical results and
allows us to explore the effects of activity via only two parameters, $\sigma_0$ and $\tau_p^b$. These are proportional to $f_0$ and $\tau_p$, although the scales differ due to coarse-graining. We find below (see Sec. \ref{MCTcomparison}) that reasonable values of relaxation time fit parameters are obtained by relating the scales of the coarse-grained model to the parameters in particle-based simulations via $\sigma_0 = 10^{-5} f_0$ and $\tau_p^b = 10^3 \tau_p$.

We first present the results for the relaxation dynamics. We fix $T=0.015$ and study the dynamics as a function of the activity parameters $\sigma_0$ and $\tau_p^b$. To characterize the dynamics, we compute the persistence function, $P(t)=\langle\tilde{P}(t)\rangle= \langle \sum_i p_i(t)/N \rangle$, where $p_i(t)=1$ if the site has not yielded within time $t$ and $0$ otherwise. We start the simulations with random initial configurations, with each $\sigma_i$ drawn from a normal distribution with zero mean and unit standard deviation. Before taking data, we evolve the systems for at least $10^4$ time steps and the measurements are taken at steady state.
We show $P(t)$ for different $\sigma_0$ and $\tau_p^b=1$ in Fig. \ref{fig-1}(a) and for varying $\tau_p^b$ and $\sigma_0=5$ in Fig. \ref{fig-1}(b). We extract the relaxation time $\tau$ from the criterion $P(\tau)=1/2$. Figure \ref{fig-1}(c) shows that for fixed $\sigma_0$, $\tau$ decreases as $\tau_p^b$ increases and then saturates, and Fig. \ref{fig-1}(d) shows that $\tau$ decreases with increasing $\sigma_0$ at fixed $\tau_p^b$. Figures \ref{fig-1}(a) and (b) show that $P(t)$ decays faster with increasing $\sigma_0$ and $\tau_p^b$; this is consistent with the ABP type of activity, where increasing $\tau_p^b$ or $\sigma_0$ fluidizes the system. Fluidization leads to decreasing $\tau$ with increasing activity, and the qualitative trends of $\tau$ with varying $\tau_p^b$ and $\sigma_0$ are similar to simulation results of active glassy systems \cite{mandal2016,nandi2017,nandi2018,paul2023}.

\begin{figure*}
	\centering
	\includegraphics[width=16.6cm]{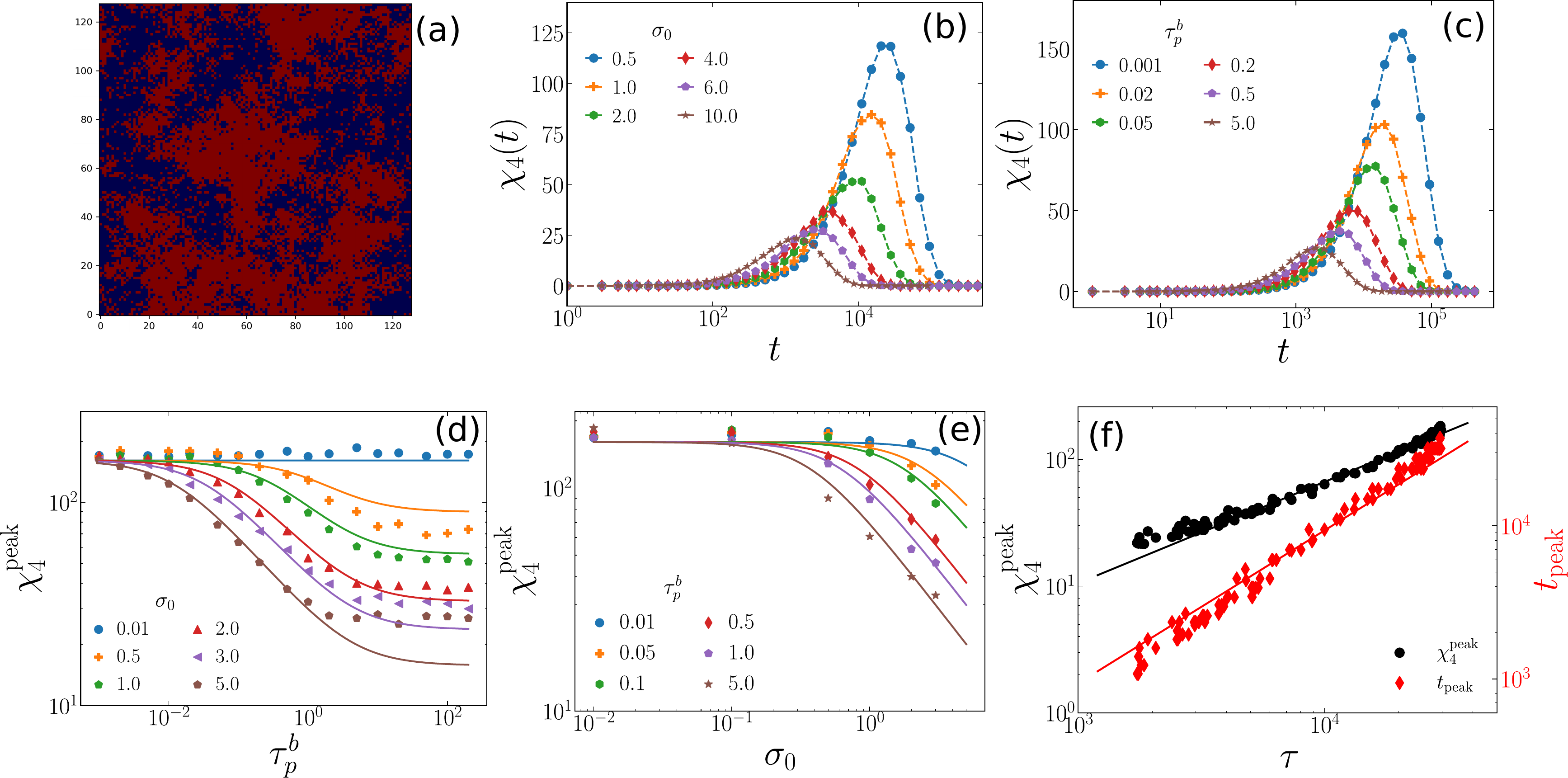}
	\caption{The behavior of the four-point correlation function. (a) Illustration of the dynamical heterogeneity at time $\tau$. We mark the individual sites red if they have yielded within time $\tau$ and blue if they have not. The parameters for the system are $T=0.015$, $\sigma_0=1$, $\tau_p^b=1$, and $L=128$. (b) Evolution of $\chi_4(t)$ for different values of $\sigma_0$ and fixed $\tau_p^b=1$. (c) Evolution of $\chi_4(t)$ for various $\tau_p^b$ and $\sigma_0=5$. (d) $\chi_4^\text{peak}$ as a function of $\tau_p^b$ for several $\sigma_0$. (e) $\chi_4^\text{peak}$ as a function of $\sigma_0$ for various $\tau_p^b$. For (d) and (e), symbols represent the simulation data, and the dashed lines are active IMCT predictions, Eq.~(\ref{mctchi4}), with $\gamma = 0.4$, $D = 160.77$, $B =3.3$, and $C = 0.25$. (f) The right y-axis shows $t_\text{peak}$. It is reasonably proportional to $\tau$; the red line is the linear fit $t_\text{peak}\simeq0.94 \tau$. The left y-axis shows $\cp$. This varies sublinearly with $\tau$ as shown by the red line, a power law with exponent $\zeta=0.8$.}
	\label{dhmct}
\end{figure*}

\subsection{Comparison to active mode-coupling theory}
\label{MCTcomparison}
We now analyze the data using active mode-coupling theory (MCT) \cite{nandi2017,activematterreview} for a more quantitative comparison of $\tau$ with known results. Within the theory, all effects of activity enter via a global effective temperature, $\Teff^G$, that is different from the local effective temperature $\Teff$.
Active MCT predicts the scaling of the relaxation time
\begin{equation}\label{mcttau}
	\tau =\tau_0 (\Teff^G - \Tmct)^{-\mu},
\end{equation}
where $\Tmct$ is the mode-coupling critical point, $\tau_0$ is a high-$T$ time scale, and $\mu$ is the MCT exponent. For the global effective temperature we write, following Refs.~\cite{benisaac2015,nandi2017,nandi2018} $\Teff^G = T + \frac{\tilde{B} \sigma_0^2 \tau_p^b}{1+C\tau_p^b}$. Here the factors $\tilde{B}$ and $C$ also account for the fact that $f_0$ and $\tau_p$ are scaled relative to their particle-level counterparts. Using this expression of $\Teff^G$ in \eqref{mcttau} gives 
\begin{equation}\label{mcttaufit}
	\tau=A\left(1+\frac{B \sigma_0^2\tau_p^b}{1+C\tau_p^b}\right)^{-\mu},
\end{equation}
where $A=\tau_0/(T-\Tmct)^\mu$ and $B=\tilde{B}/(T-\Tmct)$ will depend on $T$, whereas $C$ and $\mu$ are independent of $T$. We determine these constants using one set of data, corresponding to $\sigma_0=1$, and obtain $\mu = 0.5$, $A = 166.60$, $B =3.3$, and $C = 0.25$. These values of $B$ and $C$ are similar to those in particulate simulations when we use the scale factors discussed earlier (see Sec. \ref{minmod}).
There are no other free parameters within the theory and we can compare it directly with the simulation data. The lines in Fig. \ref{fig-1}(c) show the comparison for varying $\tau_p^b$ for several $\sigma_0$, and those in Fig. \ref{fig-1}(d) show the comparison with varying $\sigma_0$ for different fixed $\tau_p^b$. Although the data deviate at higher activity, where the effective equilibrium approximation of active MCT breaks down, the agreement is good at lower activity. This comparison shows that the active EPM reproduces the relaxation dynamics seen in particle-level simulations of active glasses \cite{mandal2016,flenner2016,Berthier2017,berthier2019,activematterreview}.
Note that the MCT exponent, $\mu$, is much smaller than the lower limit for equilibrium systems ($\sim 1.7$), in a further indication of the limits of the effective equilibrium approach.

\subsection{Dynamical heterogeneity}
We now characterize the dynamical heterogeneity (DH) obtained from our active EPM simulations. DH has emerged as one of the most fascinating hallmarks of glassy systems \cite{berthier2011}. In addition, several recent studies have demonstrated that activity has nontrivial effects on the DH of glassy systems \cite{paul2023,matteo2022,ghoshal2020}. However, the source of this nontrivial behavior of DH when the relaxation dynamics remains roughly equilibrium-like at a suitably defined $\Teff^G$ is unclear. Qualitatively, DH refers to the simultaneous existence of regions in the system that relax on different time scales. Figure \ref{dhmct}(a) shows a typical snapshot of the system after time $\tau$, where blue color marks the sites that have yielded (fast) within this time and yellow marks the ones that have not yielded (slow). The two colors tend to form clusters, highlighting the spatially heterogeneous nature of the dynamics. 

The variance of $\tilde{P}(t)$ measures the DH. Thus, in analogy to the four-point correlation function in glasses \cite{dasgupta1991,IMCT}, we define $\chi_4(t)=N(\langle \tilde{P}^2(t)\rangle-\langle \tilde{P}(t)\rangle^2)$. 
To investigate the effects of activity on the DH, we take a constant $T=0.015$ and characterize $\chi_4(t)$  for varying activity parameters. We show $\chi_4(t)$ as a function of $t$ with varying $\sigma_0$ and $\tau_p^b=1$ in Fig. \ref{dhmct}(b), and for varying $\tau_p^b$ and $\sigma_0=5$ in Fig. \ref{dhmct}(c). One observes that $\chi_4(t)$ increases with $t$, reaches a peak at intermediate times, and then decays towards zero. The peak value $\cp$ of $\chi_4(t)$ and the time $t_\text{peak}$ when this peak is reached both decrease as $\sigma_0$ or $\tau_p^b$ increase. These are typical characteristics of $\chi_4(t)$ in active glasses with ABP activity.

The value of $\cp$ gives the correlation volume of dynamically correlated regions. We first characterize the behavior of $\cp$ with varying activity. We show $\cp$ as a function of $\tau_p^b$ at constant $\sigma_0$ in Fig. \ref{dhmct}(d) and as a function of $\sigma_0$ at fixed $\tau_p^b$ in Fig. \ref{dhmct}(e). The qualitative trends of $\chi_4(t)$ and $\cp$ are similar to those in particle-level simulations \cite{flenner2016,paul2023,kolya2024}. The recently developed inhomogeneous mode-coupling theory (IMCT) for active systems predicts a power law behavior of $\cp$ \cite{paul2023,kolya2024}
\begin{equation}\label{mctchi4}
	\chi_4^\text{peak}=D\left(1+\frac{Bf_0^2\tau_p^b}{1+C\tau_p^b}\right)^{-\gamma},
\end{equation}
where $\gamma$ and $D=\chi_0/(T-\Tmct)^\gamma$ are two constants and $B$ and $C$ are the same as in Eq.~(\ref{mcttaufit}); $\chi_0$ is related to the high-$T$ value of $\chi_4^\text{peak}$. As before, we can use one data set, corresponding to $\sigma_0=1$, to obtain the new constants $D=160.77$ and $\gamma\simeq0.4$. We emphasize that once these parameters are fixed, there are no other free parameters in the theory. Figures \ref{dhmct}(d) and (e) show the comparison of the IMCT predictions (lines) with the simulation data (symbols). Although they deviate at higher activity, they agree well at lower activity. This agreement shows that our minimal active EPM also captures the basic phenomenology of DH in active glasses. We will discuss the specific mechanism that controls DH in active glasses in Sec. \ref{dhmechanism}.

The time $t_\text{peak}$ can be interpreted as a decorrelation time for the dynamics and hence gives another measure of relaxation time. Therefore, we expect $t_\text{peak}\propto \tau$; Fig. \ref{dhmct}(f) shows that this expectation holds, and we obtain $t_\text{peak}\simeq 0.94 \tau$. Furthermore, since IMCT is a critical theory, its underlying physical assumption is that a diverging correlation length, and hence a corresponding correlation volume $\cp$, govern the divergence of $\tau$. IMCT predicts a power law for $\cp$ as a function of $\tau$: $\cp\sim \tau^\zeta$. Using Eqs.~(\ref{mcttaufit}) and (\ref{mctchi4}), we obtain $\zeta=\gamma/\mu$. Figure \ref{dhmct}(f) shows $\cp$ as a function of $\tau$ in the log-log plot; the line is a power law fit, $\cp=a\tau^\zeta$ with $a=0.042$ and $\zeta\simeq 0.8$. The fitted value of $\zeta=\gamma/\mu$ is consistent with the values of $\mu$ and $\gamma$.

\begin{figure*}
	\includegraphics[width=15.6cm]{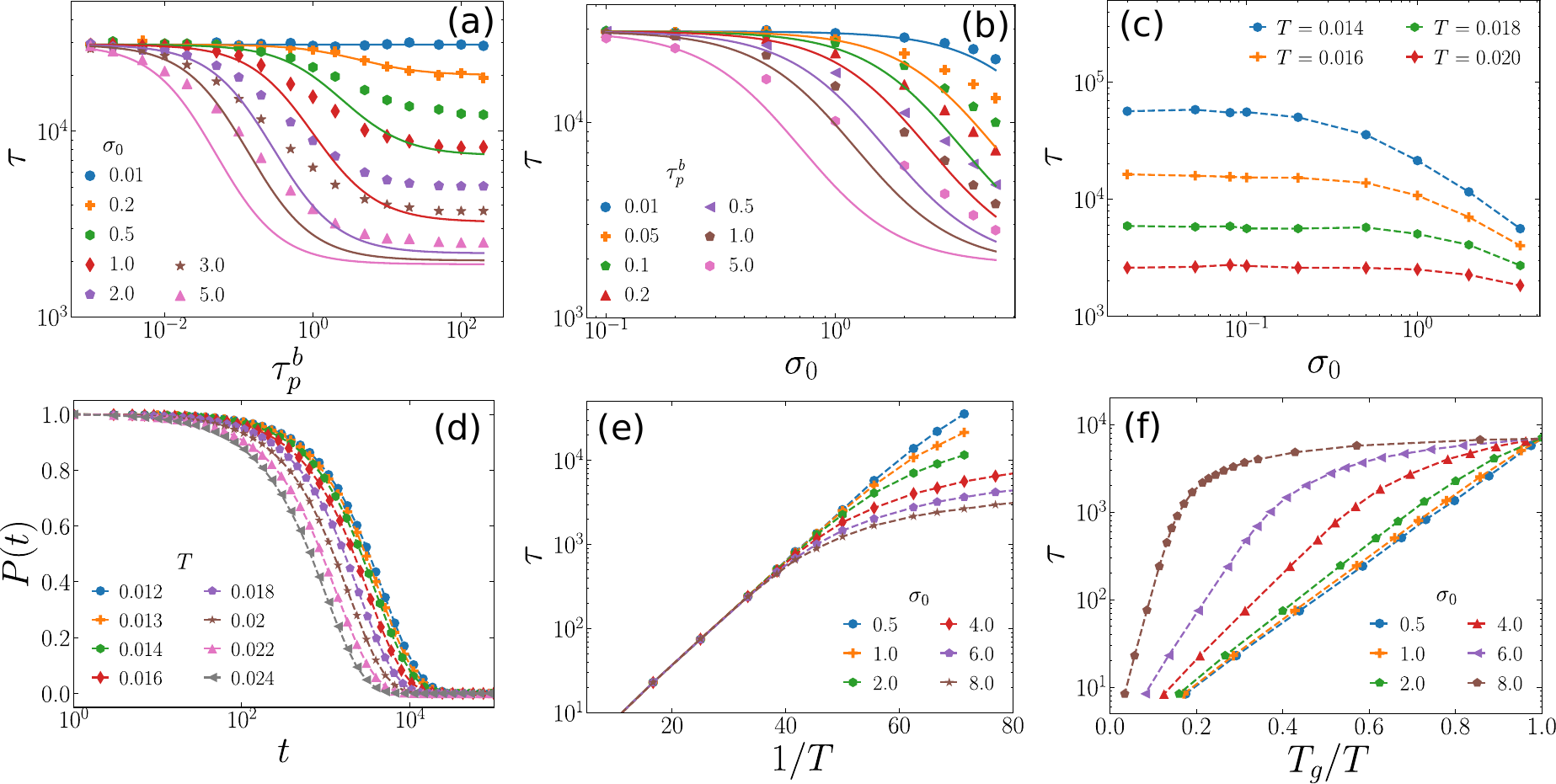}
	\caption{Relaxation dynamics analysis via active RFOT theory. (a) The relaxation time, $\tau$, as a function of $\tau_p^b$ at several values of $\sigma_0$. The lines are RFOT theory comparisons with the simulation data (symbols). (b) $\tau$ as a function of $\sigma_0$ for different values of $\tau_p^b$. The lines represent the RFOT theory; symbols show the data. (c) $\tau$ as a function of $\sigma_0$ shows active thinning behavior at different $T$. (d) Persistence curves $P(t)$ at fixed $\sigma_0=8$ and $\tau_p^b=1$ approach each other as $T$ becomes very small. (e) This overlap of $P(t)$ leads to a deviation from Arrhenius behavior at low $T$, which becomes apparent when we plot $\tau$ as a function of $1/T$: Activity dominates the relaxation dynamics at low $T$. (f) Angell plot representation of $\tau$ (see text for definitino of $T_g$): sub-Arrhenius behaviour is visible from the fact that the  
		curves like above the Arrhenius line. We have used $\tau_p^b=1$ for (c)-(f).}
	\label{rfotfit_angell}
\end{figure*}

\subsection{Comparison with active random first-order transition theory}
Another popular theory of glasses, the random first-order transition (RFOT) theory, has also been extended to active glasses \cite{nandi2018,mandal2022,sadhukhan2024}. The active RFOT theory agrees well with the existing simulation and experimental results. Therefore, we also compare the relaxation dynamics, namely the behavior of $\tau$ with varying activity, with the active RFOT theory. 
The extension of RFOT theory for active glasses treats activity as a perturbation around the passive properties,
assuming an effective equilibrium scenario \cite{nandi2018}. The dominant contribution to glassy properties comes from the vanishing of the configurational entropy at the Kauzmann transition point, where $\tau$ diverges. Therefore, Ref. \cite{nandi2018} ignored the activity modification to the surface reconfiguration energy and considered it only for the configurational entropy. Thus, we obtain the active RFOT theory for $\tau$ as
\begin{equation}\label{rfottauoriginal}
\ln\tau=\ln\tau_0+\frac{\tilde{E}}{T-T_K+\frac{\tilde{k}_1\sigma_0^2\tau_p^b}{1+k_2 \tau_p^b}},
\end{equation}
where $\tilde{E}$, $\tilde{k}_1$, and $k_2$ are constants, and $\tau_0$ is the high temperature value of $\tau$. $T_K$ is the Kauzmann temperature. For constant $T$, we can write Eq. (\ref{rfottauoriginal}) as
\begin{equation}\label{rfottau}
	\ln\tau=\ln\tau_0+\frac{E}{1+\frac{{k}_1\sigma_0^2\tau_p^b}{1+k_2\tau_p^b}},
\end{equation}
where we have redefined $E=\tilde{E}/(T-T_K)$ and $k_1=\tilde{k}_1/(T-T_K)$. Similar to the comparison with active MCT, we can obtain the constants by fitting Eq. (\ref{rfottau}) with one set of simulation data; there are then no other free parameters within the theory. 
Figures \ref{rfotfit_angell}(a) and (b) show the comparison of the active RFOT theory predictions (lines) with the data (symbols) as functions of $\tau_p^b$ and $\sigma_0$, respectively, with the constants $E=2.75$, $k_1=0.8$, $k_2=0.2$ and $\ln\tau_0=7.534$. The agreement with the theory is poorer than for active MCT, although the qualitative trends are similar. As we discuss below, the reason behind the poorer agreement with the RFOT theory is two-fold: the dynamics is sub-Arrhenius, and the minimal model breaks down at very low $T$, where RFOT theory would normally be expected to work better.

We next present results for the relaxation dynamics at varying $T$. Figure \ref{rfotfit_angell}(c) shows $\tau$ as a function of $\sigma_0$ for various $T$: when $\sigma_0$ is small, $\tau$ has similar behavior as for a thermal EPM, while when $\sigma_0$ becomes sufficiently large, $\tau$ decreases as $\sigma_0$ increases. This behavior is known as active thinning, and many simulations have shown it for glassy dynamics in particle systems \cite{flenner2016,mandal2016,paul2023}. Figure \ref{rfotfit_angell}(d) shows $P(t)$ as a function of $t$ for varying $T$ at constant $\sigma_0$ and $\tau_p^b$. As $T$ decreases, activity dominates the relaxation dynamics, and the $P(t)$ curves for various $T$ almost overlap. The plot of $\ln\tau$ as a function of $1/T$ is a straight line for the equilibrium model, which shows an Arrhenius behavior \cite{ozawa2023}. With increasing $\sigma_0$, the curves progressively deviate from this straight line as $T$ decreases or, equivalently, $1/T$ increases (Fig. \ref{rfotfit_angell}e). The deviations from Arrhenius behaviour are more visible in an Angell plot representation \cite{angell1991,angell1995} of $\tau$ as a function of $T_g/T$, where we have defined $T_g$ as the temperature where $\tau=7 \times 10^3$ (Fig. \ref{rfotfit_angell}f). 
The active EPM shows sub-Arrhenius relaxation, with the curves falling above the Arrhenius line in Fig. \ref{rfotfit_angell}(f).  Several works have shown that activity decreases fragility \cite{mandal2016,nandi2018}. Although the quantitative behavior will change if we include other consequences of activity (see Sec. \ref{indivele} below), the qualitative behavior should remain the same. Thus, the minimal active EPM already captures this aspect of decreasing fragility with increasing activity in the ABP model.

\begin{figure}
	\includegraphics[width=8.5cm]{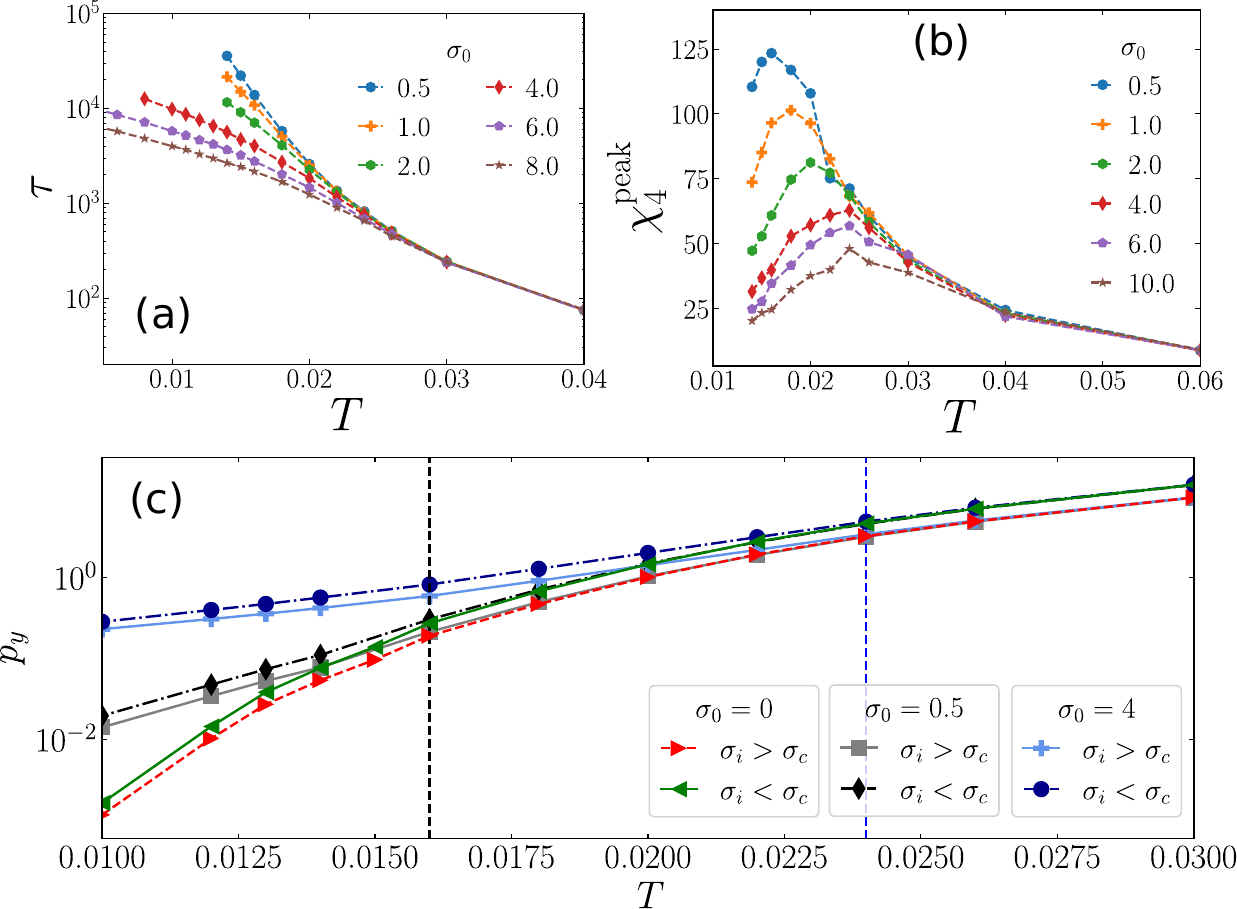}
	\caption{Temperature dependences within the minimal model. (a) Relaxation time $\tau$ as a function of $T$ for various $\sigma_0$. $\tau$ deviates from the exponential trend and saturates once $T$ becomes very small. (b) $\cp$ shows non-monotonicity at lower $T$ within the minimal model. These results suggest the breakdown of the model at very low $T$. (c) The number of yielding events, $p_y$, per unit time for direct and thermal yields for the equilibrium and active systems, with $\sigma_0=0$ and $\sigma_0\neq 0$, respectively. The dashed lines indicate the time corresponding to the peak position of $\cp$ in (b).}
	\label{limitation}
\end{figure}

\subsection{Limitations of the minimal model}
\label{limitationminmod}

We now show that the minimal model breaks down at very low $T$, where activity becomes dominant.
In our minimal model, we have ignored $T_a$ in the activated yielding crossing with local effective temperature $\Teff=T+T_a$. Figure \ref{limitation}(a) shows $\tau$ as a function of $T$. When $T$ is extremely low, relaxation continues via the active yielding, and $\tau$ saturates to a $\sigma_0$-dependent value.
This behavior results in a deviation of $\tau$ from the expected glassy behavior and shows that the minimal model fails at very low $T$. In this regime, the other elements of the active EPM, such as the persistence of $\psi$, will also be important.

On the other hand, the behavior of $\cp$ as a function of $T$ at constant activity exhibits a non-monotonic behavior within the minimal model: Figure \ref{limitation}(b) shows that $\cp$ initially increases with decreasing $T$, but then starts decreasing at very low $T$. To investigate the mechanism behind this non-monotonic behavior of $\cp$, we have studied the two distinct types of relaxations [see Fig. \ref{limitation}(c)] for $\sigma_i>\sigma_c$ (``direct yielding'') and $\sigma_i<\sigma_c$ (``thermal yielding'') for the equilibrium and the minimal active systems.
Figure \ref{limitation}(c) shows the yielding events per unit time, $p_y$, for the direct and thermal yieldings for three different $\sigma_0$. The behavior of $p_y$ for non-zero $\sigma_0$ deviates from that of $\sigma_0=0$ 
although the rates of direct and thermal yields remain nearly proportional at all $T$. This deviation shows that at very low $T$, relaxation is dominated by the active stress rate pushing local stresses close to the critical value, $\sigma_c$. As this process is independent of $T$, we see a saturation of $\tau$ at low $T$. The non-monotonic behavior of $\cp$ appears approximately when $p_y$ for the active system deviates (indicated by the dashed lines) from the equilibrium behavior. Physically, we expect $\cp$ to continue to increase as $T\to 0$ when temperature enters through $\Teff$ consistently with all three activity parameters. We therefore now briefly examine the other effects of activity within the active EPM that will be crucial for the relaxation dynamics at low $T$.


\begin{figure}
	\centering	
	\includegraphics[width=8.6cm]{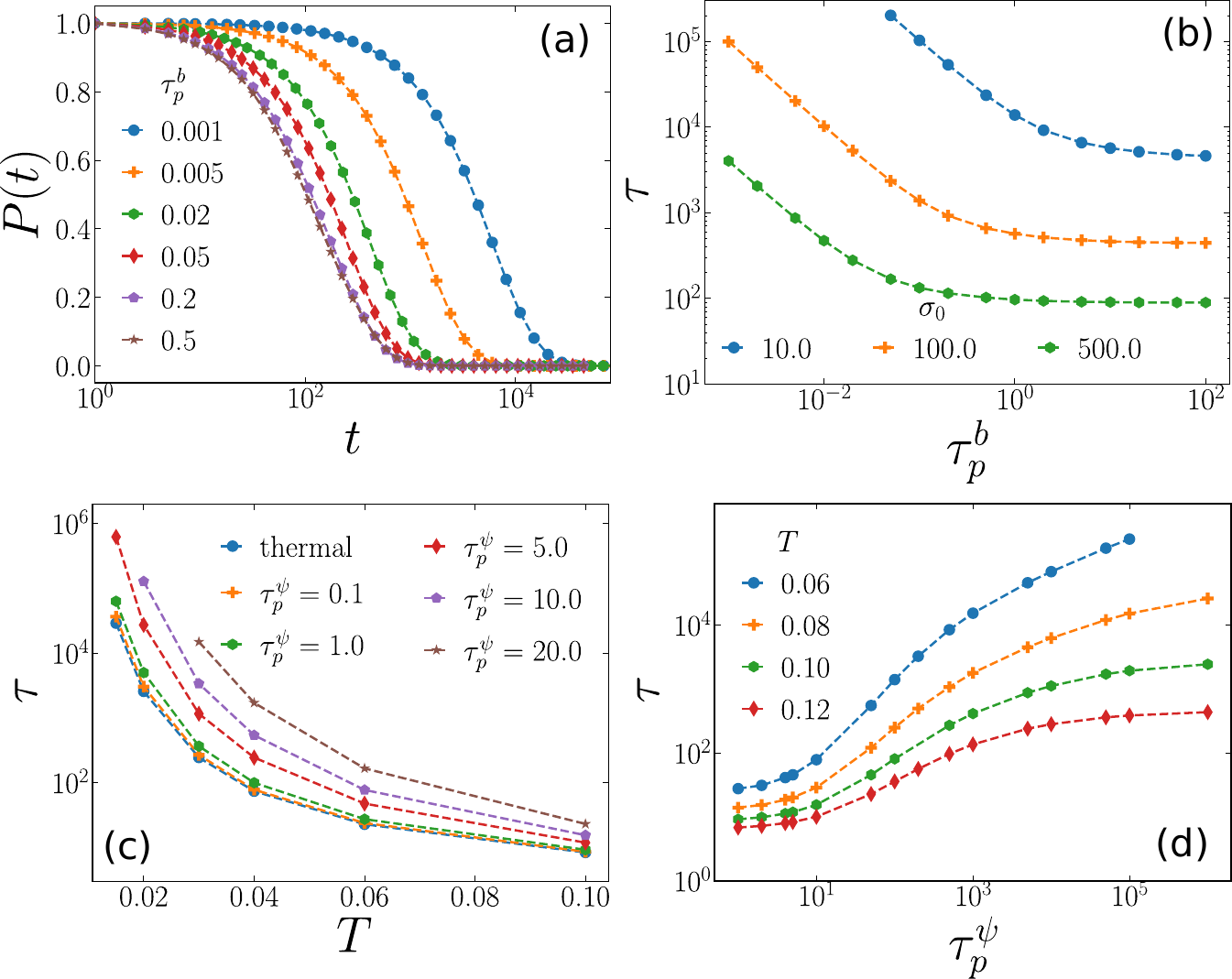}
	\caption{Effects of persistence in yielding directions $\psi_i$. (a,b) Extreme case where the $\psi_i$ are quenched such that $\tau_p^\psi=\infty$, while thermal yielding is turned off 
($\Teff=0$) so that only the active stress rate can cause yielding. (a) Decay of persistence function $P(t)$ for different values of $\tau_p^b$ and fixed $\sigma_0=500$. (b)  Relaxation time $\tau$ as a function of $\tau_p^b$ for various values of $\sigma_0$. Note that the scale of $\sigma_0$ is much higher in these two figures compared to the other figures as we have taken $\Teff=0$. (c,d) Effects of yielding direction persistence with only thermal yielding ($\sigma_0=0$, $\Teff=T>0$). (c) $\tau$ decreases with increasing $T$. (d) $\tau$ as a function of $\tau_p^\psi$ is nearly constant when $\tau_p^\psi$ is small, increases as $\tau_p^\psi$ grows and then saturates to a constant. }
	\label{individualelements}
\end{figure}

\begin{figure*}
	\includegraphics[width=16.6cm]{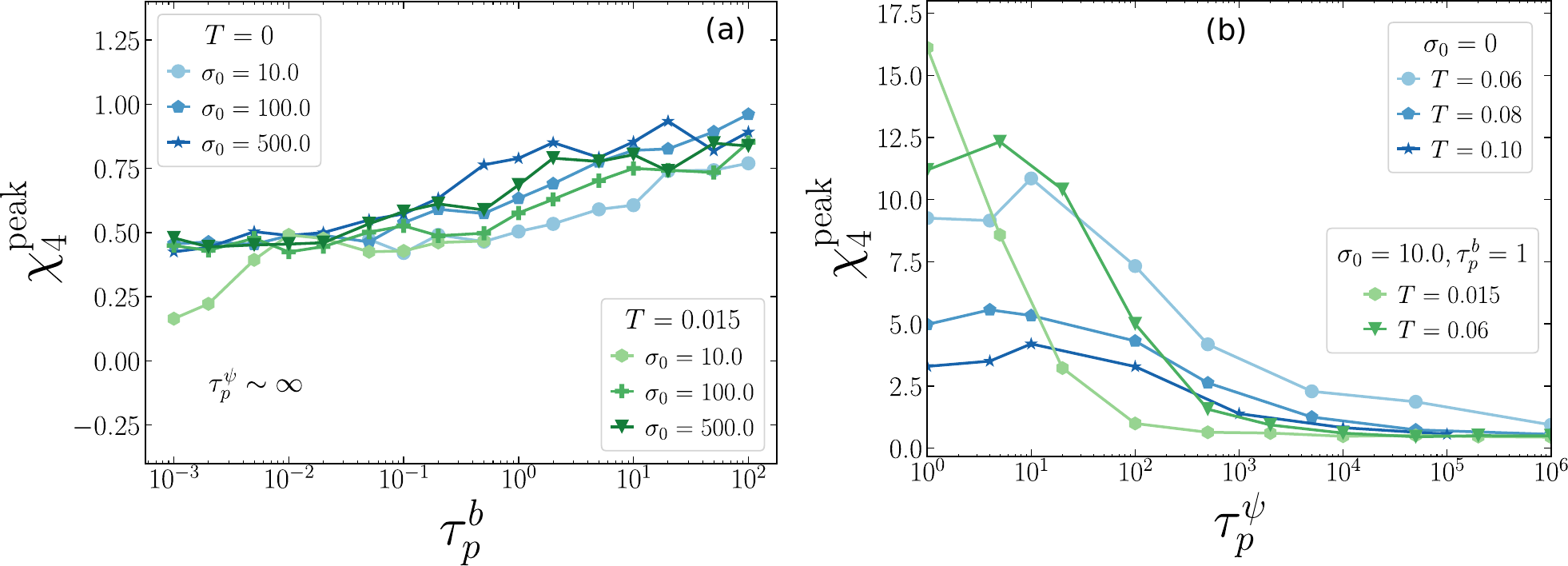}
	\caption{Persistence of yielding direction governs DH. (a) When we set $\tau_p^\psi=\infty$, $\cp$ remains small and nearly constant irrespective of the other parameter values. (b) For a given set of parameters, $\cp$ increases as $\tau_p^\psi$ decreases.}
	\label{chi4growth}
\end{figure*}

\subsection{Effects of the individual  active EPM features}
\label{indivele}

As discussed in the last section, understanding the effect of activity at very low $T$ requires going beyond the minimal model and incorporating the other model features. We now briefly discuss how these features individual affect the dynamics within our active EPM. 

{\em Relaxation at infinite $\tau_p^\psi$ via active stress rate}: 
We first discuss the effects of persistence in the yielding directions $\psi_i$, specifically, the extreme case where the $\psi_i$ are assigned randomly and then quenched, corresponding to $\tau_p^\psi=\infty$. To isolate the effects of this persistence we set $\Teff=0$ so that yielding can only be direct, i.e., take place at $\sigma_i>\sigma_c$. Figure \ref{individualelements}(a) shows $P(t)$ for various $\tau_p^b$ with $\sigma_0 = 0.005$. The decay of $P(t)$ becomes faster as $\tau_p^b$ increases: at constant $\sigma_0$, higher active stress then builds up at a site so that yielding is enhanced. Figure \ref{individualelements}(b) shows $\tau$ as a function of $\tau_p^b$, showing similar behavior also for other $\sigma_0$; increasing $\sigma_0$ accelerates yielding as expected. Looking in more detail at the $\tau_p^b$-dependence, 
$\tau$ diverges as $\tau_p^b \to0$ and saturates to a $\sigma_0$-dependent value as $\tau_p^b\to\infty$.
This relaxation dynamics via an active stress rate is novel to active systems. The results in Figs. \ref{individualelements} (a) and (b) are similar to the relaxation dynamics of athermal ABP systems \cite{matteo2022,nandi2017,activematterreview}. This similarity suggests that $b_i$ is one of the main mechanisms of relaxation dynamics in active systems.

{\em Relaxation at finite $\tau_p^\psi$ via thermal yielding}:
We next study what effects persistence of yielding directions has on relaxation caused by thermally activated yielding at temperature $T$. Again we seek to isolate individual model features and so set $\sigma_0=0$, turning off the active stress rate. We show $\tau$ as a function of $T$ for various values of $\tau_p^\psi$ in Fig. \ref{individualelements}(c). Small $\tau_p^\psi$ produce effectively thermal behavior, with the same dependence on temperature $T$ as found by Ozawa and Biroli \cite{ozawa2023}, while for larger $\tau_p^\psi$, the behavior changes from the thermal one.
Figure \ref{individualelements}(d) shows the dependence on $\tau_p^\psi$ at different constant $T$. $\tau$ grows as $\tau_p^\psi$ increases and saturates to a $T$-dependent value at very large $\tau_p^\psi$, when the $\psi_i$ become effectively quenched. Thus, apart from the effect of $\sigma_0$ and $\tau_p^b$, $\tau_p^\psi$ also crucially affects the relaxation dynamics in active systems. We next show that $\tau_p^\psi$ primarily controls DH in active glasses.


\subsection{Persistence time of yielding direction controls dynamic heterogeneity in active glasses}
\label{dhmechanism}

Finally, we demonstrate that the persistent random walk of $\psi$ governs the DH in active systems. We first set $\tau_p^\psi=\infty$ and vary $T$ and $\sigma_0$. Figure \ref{chi4growth}(a) shows $\cp$ as a function of $\tau_p^b$ for $T=0$ and $T=0.015$, for three values of $\sigma_0$ at each $T$. $\cp$ remains very small and nearly constant when $\tau_p^\psi=\infty$, independently of the other parameter values. Thus, the stochastic dynamics of $\psi$ is crucial for generating  DH. We now show that the persistence of the random walk controls the degree of DH. Figure \ref{chi4growth}(b) shows $\cp$ as a function of $\tau_p^\psi$ for $\sigma_0=0$ and ($\sigma_0=10.0$, $\tau_p^b=1$), both combined with different $T$. Of course, there are inherent inconsistencies in these various parameters as non-zero $\tau_p^\psi$ requires a non-zero active temperature and this shows up as a non-monotonic trend in $\cp$ discussed in Sec. \ref{limitationminmod}. However, the main trend is unambiguous: as $\tau_p^\psi$ decreases, $\cp$ and hence DH increases. Overall, we see that two distinct effects control the relaxation dynamics and DH in active systems: whereas the active yielding primarily determines $\tau$, the persistence of the yielding direction governs the DH.

\section{Discussion}
To conclude, we have developed an elastoplastic model for active glasses and demonstrated that it agrees with the known results of active glassy dynamics in particle systems. Many experiments in the last couple of decades have shown that the effects of activity on the glassy dynamics are crucial for many biologically significant processes. For example, during embryogenesis and cancer progression, the solid-like epithelial state fluidizes during the epithelial-to-mesenchymal transition (EMT), when the mesenchymal cells acquire a motile nature and become active \cite{thiery2002,thiery2006,mohit2015,sadhukhan2024}. A similar activity-mediated glass transition exists in several other systems \cite{nishizawa2017,angelini2011,park2015,fabry2001,deng2006,tambe2011}. However, the nonequilibrium nature of the problem and the inherent complexities of biological systems make them very challenging for a detailed theoretical understanding \cite{activematterreview,janssen2019}. Simulations of particle systems can and have provided crucial insights but remain limited by the large computation time requirements. Diverse forms of activity further complicate the matter \cite{ramaswamy2010,marchetti2013,jacques2015}. Our mesoscopic approach via the active EPM provides a novel and complementary way of investigating the glassy dynamics in these systems.

Recent simulation results show that whereas the relaxation dynamics remains equilibrium-like, activity has nontrivial effects on the DH \cite{berthier2019,Berthier2017,paul2023,activematterreview,kolya2024,matteo2022,zheng2024,mandal2020PRL,mandal2021}. Mode-coupling theory traces this to two distinct mechanisms for the relaxation dynamics and DH in active glasses \cite{kolya2024}. Our active EPM posits a very similar scenario. In addition, the active EPM points to the individual mechanisms: whereas the active yielding -- driven by a fluctuating active stress rate -- primarily controls the relaxation dynamics, the persistence of the yielding direction governs the DH in active systems. This distinct behavior is a novel aspect of active systems, where persistence time differentiates active driving from thermal noise. Thus, the relaxation dynamics and the DH in active systems can become uncorrelated in some regimes of the parameter space.

Several studies have shown the existence of a long-range effective interaction in active systems \cite{caprini2020,caprini2020b,chate2006}. Since EPMs naturally include such interactions via the stress redistribution propagator, these models may work even better for active systems. One crucial advantage of the elastoplastic approach is that we can separately analyze various aspects of activity, and this can provide deeper insights. 
In this work, we have studied a minimal version of an active EPM and showed that including the yielding via active stress build-up within a thermal EPM already captures most aspects of active glasses within a reasonable range of parameters. We have also investigated the detailed effects of a further element of the active EPM, the persistence of the yield directions $\psi_i$. It will be interesting to extend this investigation to include the active contribution $T_a$ to the barrier crossing during yielding, which is also determined by the underlying activity parameters. Since activity can have many different forms, several variants of the active EPM are possible. In the future, we will look at other versions of an active EPM and explore the role of different forms of activity. Specifically, the current model is a scalar version of the EPM, whereas stress is a tensor. Therefore, the tensorial EPM \cite{tahaei2023} will be more appropriate to study different aspects of activity. 
In addition, compared to particulate simulations, EPMs allow studying a relatively lager system size and longer time scale that will be advantageous for the investigation of glassy properties.
Furthermore, we have considered the active stress rates to be spatially uncorrelated, while they could certainly be correlated if they are e.g.\ generated by local body forces as in ABPs. Including spatial correlations of active stresses is therefore an interesting extension that we will also explore in future.
Summarizing, even a minimal version of an active EPM studied in this work already shows promising results demonstrating the specific distinct mechanisms of relaxation dynamics and DH, and can act as a solid base from which to study future extensions. 

\appendix

\begin{figure}
	\includegraphics[width=8.4cm]{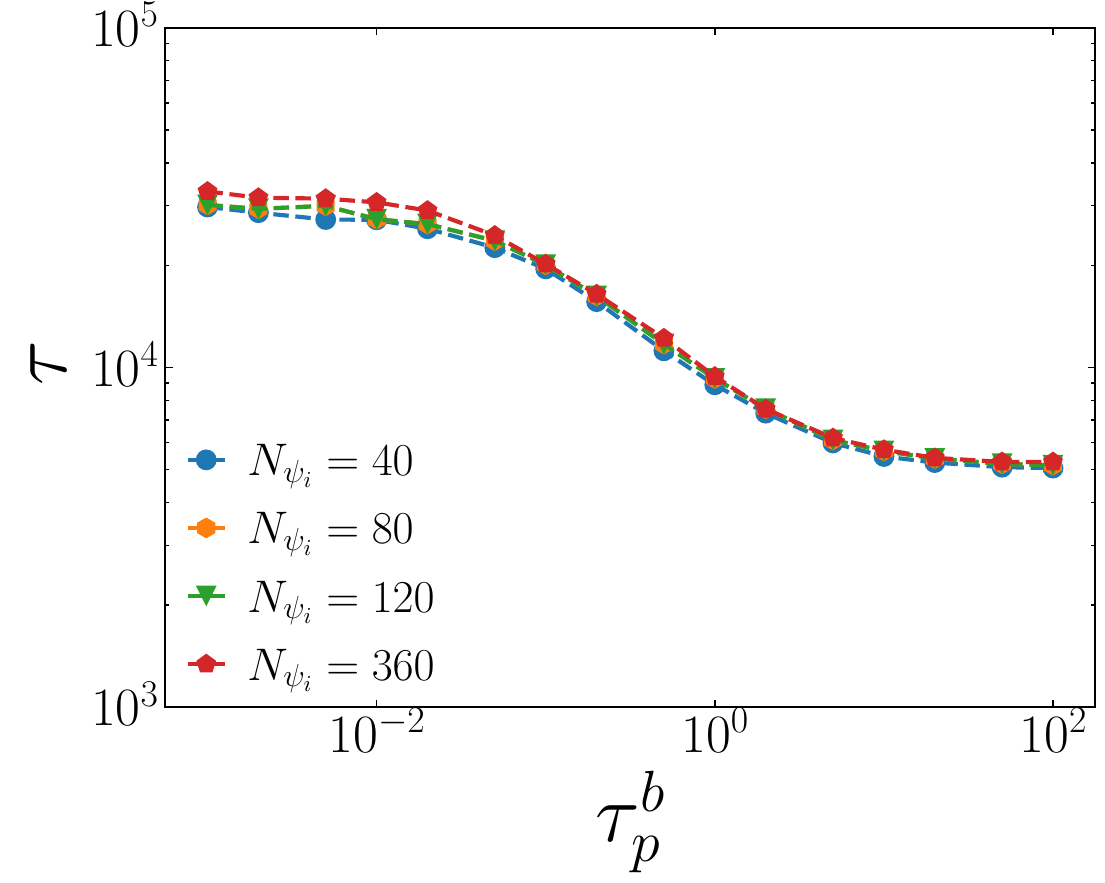}
	\caption{Effects of the discretization of $\psi_i$. Relaxation time $\tau$ as a function of $\tau_p^b$ for different values of $N_{\psi_i}$. We have used $\sigma_0 = 2 $ and $T= 0.015$. }
	\label{fig-sm1}  
\end{figure}

\section{Technical details on the implementation of the active EPM} 
The stress propagator $\Gij$ as
\begin{equation}
	\Gij = \frac{\cos[4(\theta_i-\psi_i)]}{\pi r_{ij}^2},
\end{equation}
where $\psi_i$ is the direction of yielding while $\theta_i$ gives the direction (polar angle) of site $j$ as seen from site $i$, and $r_{ij}$ is the distance between the two sites. The computation of this Eshelby kernel at every simulation step where a yield event occurs is costly. To alleviate this difficulty, we have discretized angles in the step of $\pi/2N_{\psi_i}$ and used predetermined values of $\Gij$ for the system. Figure \ref{fig-sm1} shows $\tau$ as a function of $\tau_p^b$ for our active EPM with varying discretizations. For the results presented in the paper we have used $L=64$ unless stated otherwise, and $N_{\psi_i}=40$.

\subsection*{Acknowledgment}
TG and SKN acknowledge the support of the Department of Atomic Energy, Government of India, under Project Identification No. RTI 4007.
PS and SKN would like to thank the Erwin Schr{\"{o}}dinger International Institute for Mathematics and Physics (ESI), University of Vienna (Austria) for the invitation to participate in the Thematic Program “Linking Microscopic Processes to the Macroscopic Rheological Properties in Inert and Living Soft Materials" in 2024 where part of this work has been carried out and for the support given. 
 SKN thanks SERB for grant via SRG/2021/002014.

\bibliography{refs_activeEPM.bib}

\begin{thebibliography}{91}%
\makeatletter
\providecommand \@ifxundefined [1]{%
 \@ifx{#1\undefined}
}%
\providecommand \@ifnum [1]{%
 \ifnum #1\expandafter \@firstoftwo
 \else \expandafter \@secondoftwo
 \fi
}%
\providecommand \@ifx [1]{%
 \ifx #1\expandafter \@firstoftwo
 \else \expandafter \@secondoftwo
 \fi
}%
\providecommand \natexlab [1]{#1}%
\providecommand \enquote  [1]{``#1''}%
\providecommand \bibnamefont  [1]{#1}%
\providecommand \bibfnamefont [1]{#1}%
\providecommand \citenamefont [1]{#1}%
\providecommand \href@noop [0]{\@secondoftwo}%
\providecommand \href [0]{\begingroup \@sanitize@url \@href}%
\providecommand \@href[1]{\@@startlink{#1}\@@href}%
\providecommand \@@href[1]{\endgroup#1\@@endlink}%
\providecommand \@sanitize@url [0]{\catcode `\\12\catcode `\$12\catcode
  `\&12\catcode `\#12\catcode `\^12\catcode `\_12\catcode `\%12\relax}%
\providecommand \@@startlink[1]{}%
\providecommand \@@endlink[0]{}%
\providecommand \url  [0]{\begingroup\@sanitize@url \@url }%
\providecommand \@url [1]{\endgroup\@href {#1}{\urlprefix }}%
\providecommand \urlprefix  [0]{URL }%
\providecommand \Eprint [0]{\href }%
\providecommand \doibase [0]{http://dx.doi.org/}%
\providecommand \selectlanguage [0]{\@gobble}%
\providecommand \bibinfo  [0]{\@secondoftwo}%
\providecommand \bibfield  [0]{\@secondoftwo}%
\providecommand \translation [1]{[#1]}%
\providecommand \BibitemOpen [0]{}%
\providecommand \bibitemStop [0]{}%
\providecommand \bibitemNoStop [0]{.\EOS\space}%
\providecommand \EOS [0]{\spacefactor3000\relax}%
\providecommand \BibitemShut  [1]{\csname bibitem#1\endcsname}%
\let\auto@bib@innerbib\@empty
\bibitem [{\citenamefont {Jawerth}\ \emph {et~al.}(2018)\citenamefont
  {Jawerth}, \citenamefont {Ijavi}, \citenamefont {Ruer}, \citenamefont {Saha},
  \citenamefont {Jahnel}, \citenamefont {Hyman}, \citenamefont {J\"ulicher},\
  and\ \citenamefont {Fischer-Friedrich}}]{jawerth2018}%
  \BibitemOpen
  \bibfield  {author} {\bibinfo {author} {\bibfnamefont {L.~M.}\ \bibnamefont
  {Jawerth}}, \bibinfo {author} {\bibfnamefont {M.}~\bibnamefont {Ijavi}},
  \bibinfo {author} {\bibfnamefont {M.}~\bibnamefont {Ruer}}, \bibinfo {author}
  {\bibfnamefont {S.}~\bibnamefont {Saha}}, \bibinfo {author} {\bibfnamefont
  {M.}~\bibnamefont {Jahnel}}, \bibinfo {author} {\bibfnamefont {A.~A.}\
  \bibnamefont {Hyman}}, \bibinfo {author} {\bibfnamefont {F.}~\bibnamefont
  {J\"ulicher}}, \ and\ \bibinfo {author} {\bibfnamefont {E.}~\bibnamefont
  {Fischer-Friedrich}},\ }\href {\doibase 10.1103/PhysRevLett.121.258101}
  {\bibfield  {journal} {\bibinfo  {journal} {Phys. Rev. Lett.}\ }\textbf
  {\bibinfo {volume} {121}},\ \bibinfo {pages} {258101} (\bibinfo {year}
  {2018})}\BibitemShut {NoStop}%
\bibitem [{\citenamefont {Jawerth}\ \emph {et~al.}(2020)\citenamefont
  {Jawerth}, \citenamefont {Fischer-Friedrich}, \citenamefont {Saha},
  \citenamefont {Wang}, \citenamefont {Franzmann}, \citenamefont {Zhang},
  \citenamefont {Sachweh}, \citenamefont {Ruer}, \citenamefont {Ijavi},
  \citenamefont {Saha} \emph {et~al.}}]{jawerth2020}%
  \BibitemOpen
  \bibfield  {author} {\bibinfo {author} {\bibfnamefont {L.}~\bibnamefont
  {Jawerth}}, \bibinfo {author} {\bibfnamefont {E.}~\bibnamefont
  {Fischer-Friedrich}}, \bibinfo {author} {\bibfnamefont {S.}~\bibnamefont
  {Saha}}, \bibinfo {author} {\bibfnamefont {J.}~\bibnamefont {Wang}}, \bibinfo
  {author} {\bibfnamefont {T.}~\bibnamefont {Franzmann}}, \bibinfo {author}
  {\bibfnamefont {X.}~\bibnamefont {Zhang}}, \bibinfo {author} {\bibfnamefont
  {J.}~\bibnamefont {Sachweh}}, \bibinfo {author} {\bibfnamefont
  {M.}~\bibnamefont {Ruer}}, \bibinfo {author} {\bibfnamefont {M.}~\bibnamefont
  {Ijavi}}, \bibinfo {author} {\bibfnamefont {S.}~\bibnamefont {Saha}},  \emph
  {et~al.},\ }\href {\doibase 10.1126/science.aaw4951} {\bibfield  {journal}
  {\bibinfo  {journal} {Science}\ }\textbf {\bibinfo {volume} {370}},\ \bibinfo
  {pages} {1317} (\bibinfo {year} {2020})}\BibitemShut {NoStop}%
\bibitem [{\citenamefont {Alshareedah}\ \emph {et~al.}(2021)\citenamefont
  {Alshareedah}, \citenamefont {Moosa}, \citenamefont {Pham}, \citenamefont
  {Potoyan},\ and\ \citenamefont {Banerjee}}]{alshareedah2021}%
  \BibitemOpen
  \bibfield  {author} {\bibinfo {author} {\bibfnamefont {I.}~\bibnamefont
  {Alshareedah}}, \bibinfo {author} {\bibfnamefont {M.~M.}\ \bibnamefont
  {Moosa}}, \bibinfo {author} {\bibfnamefont {M.}~\bibnamefont {Pham}},
  \bibinfo {author} {\bibfnamefont {D.~A.}\ \bibnamefont {Potoyan}}, \ and\
  \bibinfo {author} {\bibfnamefont {P.~R.}\ \bibnamefont {Banerjee}},\
  }\href@noop {} {\bibfield  {journal} {\bibinfo  {journal} {Nature
  communications}\ }\textbf {\bibinfo {volume} {12}},\ \bibinfo {pages} {6620}
  (\bibinfo {year} {2021})}\BibitemShut {NoStop}%
\bibitem [{\citenamefont {Fabry}\ \emph {et~al.}(2001)\citenamefont {Fabry},
  \citenamefont {Maksym}, \citenamefont {Butler}, \citenamefont {Glogauer},
  \citenamefont {Navajas},\ and\ \citenamefont {Fredberg}}]{fabry2001}%
  \BibitemOpen
  \bibfield  {author} {\bibinfo {author} {\bibfnamefont {B.}~\bibnamefont
  {Fabry}}, \bibinfo {author} {\bibfnamefont {G.~N.}\ \bibnamefont {Maksym}},
  \bibinfo {author} {\bibfnamefont {J.~P.}\ \bibnamefont {Butler}}, \bibinfo
  {author} {\bibfnamefont {M.}~\bibnamefont {Glogauer}}, \bibinfo {author}
  {\bibfnamefont {D.}~\bibnamefont {Navajas}}, \ and\ \bibinfo {author}
  {\bibfnamefont {J.~J.}\ \bibnamefont {Fredberg}},\ }\href {\doibase
  10.1103/PhysRevLett.87.148102} {\bibfield  {journal} {\bibinfo  {journal}
  {Phys. Rev. Lett.}\ }\textbf {\bibinfo {volume} {87}},\ \bibinfo {pages}
  {148102} (\bibinfo {year} {2001})}\BibitemShut {NoStop}%
\bibitem [{\citenamefont {Deng}\ \emph {et~al.}(2006)\citenamefont {Deng},
  \citenamefont {Trepat}, \citenamefont {Butler}, \citenamefont {Millet},
  \citenamefont {Morgan}, \citenamefont {Weitz},\ and\ \citenamefont
  {Fredberg}}]{deng2006}%
  \BibitemOpen
  \bibfield  {author} {\bibinfo {author} {\bibfnamefont {L.}~\bibnamefont
  {Deng}}, \bibinfo {author} {\bibfnamefont {X.}~\bibnamefont {Trepat}},
  \bibinfo {author} {\bibfnamefont {J.~P.}\ \bibnamefont {Butler}}, \bibinfo
  {author} {\bibfnamefont {E.}~\bibnamefont {Millet}}, \bibinfo {author}
  {\bibfnamefont {K.~G.}\ \bibnamefont {Morgan}}, \bibinfo {author}
  {\bibfnamefont {D.~A.}\ \bibnamefont {Weitz}}, \ and\ \bibinfo {author}
  {\bibfnamefont {J.~J.}\ \bibnamefont {Fredberg}},\ }\href {\doibase
  10.1038/nmat1685} {\bibfield  {journal} {\bibinfo  {journal} {Nat.
  Materials}\ }\textbf {\bibinfo {volume} {5}},\ \bibinfo {pages} {636}
  (\bibinfo {year} {2006})}\BibitemShut {NoStop}%
\bibitem [{\citenamefont {Bursac}\ \emph {et~al.}(2005)\citenamefont {Bursac},
  \citenamefont {Lenormand}, \citenamefont {Fabry}, \citenamefont {Oliver},
  \citenamefont {Weitz}, \citenamefont {Viasnoff}, \citenamefont {Butler},\
  and\ \citenamefont {Fredberg}}]{bursac2005}%
  \BibitemOpen
  \bibfield  {author} {\bibinfo {author} {\bibfnamefont {P.}~\bibnamefont
  {Bursac}}, \bibinfo {author} {\bibfnamefont {G.}~\bibnamefont {Lenormand}},
  \bibinfo {author} {\bibfnamefont {B.}~\bibnamefont {Fabry}}, \bibinfo
  {author} {\bibfnamefont {M.}~\bibnamefont {Oliver}}, \bibinfo {author}
  {\bibfnamefont {D.~A.}\ \bibnamefont {Weitz}}, \bibinfo {author}
  {\bibfnamefont {V.}~\bibnamefont {Viasnoff}}, \bibinfo {author}
  {\bibfnamefont {J.~P.}\ \bibnamefont {Butler}}, \ and\ \bibinfo {author}
  {\bibfnamefont {J.~J.}\ \bibnamefont {Fredberg}},\ }\href {\doibase
  10.1038/nmat1404} {\bibfield  {journal} {\bibinfo  {journal} {Nat. Mat.}\
  }\textbf {\bibinfo {volume} {4}},\ \bibinfo {pages} {557} (\bibinfo {year}
  {2005})}\BibitemShut {NoStop}%
\bibitem [{\citenamefont {Parry}\ \emph {et~al.}(2014)\citenamefont {Parry},
  \citenamefont {Surovtsev}, \citenamefont {Cabeen}, \citenamefont {O’Hern},
  \citenamefont {Dufresne},\ and\ \citenamefont {Jacobs-Wagner}}]{parry2014}%
  \BibitemOpen
  \bibfield  {author} {\bibinfo {author} {\bibfnamefont {B.}~\bibnamefont
  {Parry}}, \bibinfo {author} {\bibfnamefont {I.}~\bibnamefont {Surovtsev}},
  \bibinfo {author} {\bibfnamefont {M.}~\bibnamefont {Cabeen}}, \bibinfo
  {author} {\bibfnamefont {C.}~\bibnamefont {O’Hern}}, \bibinfo {author}
  {\bibfnamefont {E.}~\bibnamefont {Dufresne}}, \ and\ \bibinfo {author}
  {\bibfnamefont {C.}~\bibnamefont {Jacobs-Wagner}},\ }\href {\doibase
  https://doi.org/10.1016/j.cell.2013.11.028} {\bibfield  {journal} {\bibinfo
  {journal} {Cell}\ }\textbf {\bibinfo {volume} {156}},\ \bibinfo {pages} {183}
  (\bibinfo {year} {2014})}\BibitemShut {NoStop}%
\bibitem [{\citenamefont {Angelini}\ \emph {et~al.}(2011)\citenamefont
  {Angelini}, \citenamefont {Hannezo}, \citenamefont {Trepat}, \citenamefont
  {Marquez}, \citenamefont {Fredberg},\ and\ \citenamefont
  {Weitz}}]{angelini2011}%
  \BibitemOpen
  \bibfield  {author} {\bibinfo {author} {\bibfnamefont {T.~E.}\ \bibnamefont
  {Angelini}}, \bibinfo {author} {\bibfnamefont {E.}~\bibnamefont {Hannezo}},
  \bibinfo {author} {\bibfnamefont {X.}~\bibnamefont {Trepat}}, \bibinfo
  {author} {\bibfnamefont {M.}~\bibnamefont {Marquez}}, \bibinfo {author}
  {\bibfnamefont {J.~J.}\ \bibnamefont {Fredberg}}, \ and\ \bibinfo {author}
  {\bibfnamefont {D.~A.}\ \bibnamefont {Weitz}},\ }\href {\doibase
  https://doi.org/10.1073/pnas.1010059108} {\bibfield  {journal} {\bibinfo
  {journal} {Proceedings of the National Academy of Sciences}\ }\textbf
  {\bibinfo {volume} {108}},\ \bibinfo {pages} {4714} (\bibinfo {year}
  {2011})}\BibitemShut {NoStop}%
\bibitem [{\citenamefont {Park}\ \emph {et~al.}(2015)\citenamefont {Park},
  \citenamefont {Kim}, \citenamefont {Bi}, \citenamefont {Mitchel},
  \citenamefont {Qazvini} \emph {et~al.}}]{park2015}%
  \BibitemOpen
  \bibfield  {author} {\bibinfo {author} {\bibfnamefont {J.-A.}\ \bibnamefont
  {Park}}, \bibinfo {author} {\bibfnamefont {J.~H.}\ \bibnamefont {Kim}},
  \bibinfo {author} {\bibfnamefont {D.}~\bibnamefont {Bi}}, \bibinfo {author}
  {\bibfnamefont {J.~A.}\ \bibnamefont {Mitchel}}, \bibinfo {author}
  {\bibfnamefont {N.~T.}\ \bibnamefont {Qazvini}},  \emph {et~al.},\ }\href
  {\doibase 10.1038/nmat4357} {\bibfield  {journal} {\bibinfo  {journal} {Nat.
  Mat.}\ }\textbf {\bibinfo {volume} {14}},\ \bibinfo {pages} {1040} (\bibinfo
  {year} {2015})}\BibitemShut {NoStop}%
\bibitem [{\citenamefont {Atia}\ \emph {et~al.}(2018)\citenamefont {Atia},
  \citenamefont {Bi}, \citenamefont {Sharma}, \citenamefont {Mitchel},
  \citenamefont {Gweon}, \citenamefont {A.~Koehler}, \citenamefont {DeCamp},
  \citenamefont {Lan}, \citenamefont {Kim}, \citenamefont {Hirsch},
  \citenamefont {Pegoraro}, \citenamefont {Lee}, \citenamefont {Starr},
  \citenamefont {Weitz}, \citenamefont {Martin}, \citenamefont {Park} \emph
  {et~al.}}]{atia2018}%
  \BibitemOpen
  \bibfield  {author} {\bibinfo {author} {\bibfnamefont {L.}~\bibnamefont
  {Atia}}, \bibinfo {author} {\bibfnamefont {D.}~\bibnamefont {Bi}}, \bibinfo
  {author} {\bibfnamefont {Y.}~\bibnamefont {Sharma}}, \bibinfo {author}
  {\bibfnamefont {J.~A.}\ \bibnamefont {Mitchel}}, \bibinfo {author}
  {\bibfnamefont {B.}~\bibnamefont {Gweon}}, \bibinfo {author} {\bibfnamefont
  {S.}~\bibnamefont {A.~Koehler}}, \bibinfo {author} {\bibfnamefont {S.~J.}\
  \bibnamefont {DeCamp}}, \bibinfo {author} {\bibfnamefont {B.}~\bibnamefont
  {Lan}}, \bibinfo {author} {\bibfnamefont {J.~H.}\ \bibnamefont {Kim}},
  \bibinfo {author} {\bibfnamefont {R.}~\bibnamefont {Hirsch}}, \bibinfo
  {author} {\bibfnamefont {A.~F.}\ \bibnamefont {Pegoraro}}, \bibinfo {author}
  {\bibfnamefont {K.~H.}\ \bibnamefont {Lee}}, \bibinfo {author} {\bibfnamefont
  {J.~R.}\ \bibnamefont {Starr}}, \bibinfo {author} {\bibfnamefont {D.~A.}\
  \bibnamefont {Weitz}}, \bibinfo {author} {\bibfnamefont {A.~C.}\ \bibnamefont
  {Martin}}, \bibinfo {author} {\bibfnamefont {J.-A.}\ \bibnamefont {Park}},
  \emph {et~al.},\ }\href {\doibase 10.1038/s41567-018-0089-9} {\bibfield
  {journal} {\bibinfo  {journal} {Nat. Phys.}\ }\textbf {\bibinfo {volume}
  {14}},\ \bibinfo {pages} {613} (\bibinfo {year} {2018})}\BibitemShut
  {NoStop}%
\bibitem [{\citenamefont {Malinverno}\ \emph {et~al.}(2017)\citenamefont
  {Malinverno}, \citenamefont {Corallino}, \citenamefont {Giavazzi},
  \citenamefont {Bergert}, \citenamefont {Li}, \citenamefont {Leoni},
  \citenamefont {Disanza}, \citenamefont {Frittoli}, \citenamefont {Oldani},
  \citenamefont {Martini} \emph {et~al.}}]{malinverno2017}%
  \BibitemOpen
  \bibfield  {author} {\bibinfo {author} {\bibfnamefont {C.}~\bibnamefont
  {Malinverno}}, \bibinfo {author} {\bibfnamefont {S.}~\bibnamefont
  {Corallino}}, \bibinfo {author} {\bibfnamefont {F.}~\bibnamefont {Giavazzi}},
  \bibinfo {author} {\bibfnamefont {M.}~\bibnamefont {Bergert}}, \bibinfo
  {author} {\bibfnamefont {Q.}~\bibnamefont {Li}}, \bibinfo {author}
  {\bibfnamefont {M.}~\bibnamefont {Leoni}}, \bibinfo {author} {\bibfnamefont
  {A.}~\bibnamefont {Disanza}}, \bibinfo {author} {\bibfnamefont
  {E.}~\bibnamefont {Frittoli}}, \bibinfo {author} {\bibfnamefont
  {A.}~\bibnamefont {Oldani}}, \bibinfo {author} {\bibfnamefont
  {E.}~\bibnamefont {Martini}},  \emph {et~al.},\ }\href {\doibase
  https://doi.org/10.1038/nmat4848} {\bibfield  {journal} {\bibinfo  {journal}
  {Nature materials}\ }\textbf {\bibinfo {volume} {16}},\ \bibinfo {pages}
  {587} (\bibinfo {year} {2017})}\BibitemShut {NoStop}%
\bibitem [{\citenamefont {Takatori}\ and\ \citenamefont
  {Mandadapu}(2020)}]{takatori2020}%
  \BibitemOpen
  \bibfield  {author} {\bibinfo {author} {\bibfnamefont {S.~C.}\ \bibnamefont
  {Takatori}}\ and\ \bibinfo {author} {\bibfnamefont {K.~K.}\ \bibnamefont
  {Mandadapu}},\ }\href {\doibase https://doi.org/10.48550/arXiv.2003.05618}
  {\bibfield  {journal} {\bibinfo  {journal} {arXiv preprint arXiv:2003.05618}\
  } (\bibinfo {year} {2020}),\
  https://doi.org/10.48550/arXiv.2003.05618}\BibitemShut {NoStop}%
\bibitem [{\citenamefont {Lama}\ \emph {et~al.}(2024)\citenamefont {Lama},
  \citenamefont {Yamamoto}, \citenamefont {Furuta}, \citenamefont {Shimaya},\
  and\ \citenamefont {Takeuchi}}]{lama2024}%
  \BibitemOpen
  \bibfield  {author} {\bibinfo {author} {\bibfnamefont {H.}~\bibnamefont
  {Lama}}, \bibinfo {author} {\bibfnamefont {M.~J.}\ \bibnamefont {Yamamoto}},
  \bibinfo {author} {\bibfnamefont {Y.}~\bibnamefont {Furuta}}, \bibinfo
  {author} {\bibfnamefont {T.}~\bibnamefont {Shimaya}}, \ and\ \bibinfo
  {author} {\bibfnamefont {K.~A.}\ \bibnamefont {Takeuchi}},\ }\href {\doibase
  10.1093/pnasnexus/pgae238} {\bibfield  {journal} {\bibinfo  {journal} {PNAS
  Nexus}\ }\textbf {\bibinfo {volume} {3}},\ \bibinfo {pages} {pgae238}
  (\bibinfo {year} {2024})}\BibitemShut {NoStop}%
\bibitem [{\citenamefont {Berthier}\ \emph {et~al.}(2019)\citenamefont
  {Berthier}, \citenamefont {Flenner},\ and\ \citenamefont
  {Szamel}}]{berthier2019}%
  \BibitemOpen
  \bibfield  {author} {\bibinfo {author} {\bibfnamefont {L.}~\bibnamefont
  {Berthier}}, \bibinfo {author} {\bibfnamefont {E.}~\bibnamefont {Flenner}}, \
  and\ \bibinfo {author} {\bibfnamefont {G.}~\bibnamefont {Szamel}},\ }\href
  {\doibase 10.1063/1.5093240} {\bibfield  {journal} {\bibinfo  {journal} {J.
  Chem. Phys.}\ }\textbf {\bibinfo {volume} {150}},\ \bibinfo {pages} {200901}
  (\bibinfo {year} {2019})}\BibitemShut {NoStop}%
\bibitem [{\citenamefont {Janssen}(2019)}]{janssen2019}%
  \BibitemOpen
  \bibfield  {author} {\bibinfo {author} {\bibfnamefont {L.~M.~C.}\
  \bibnamefont {Janssen}},\ }\href {\doibase 10.1088/1361-648X/ab3e90}
  {\bibfield  {journal} {\bibinfo  {journal} {Journal of Physics: Condensed
  Matter}\ }\textbf {\bibinfo {volume} {31}},\ \bibinfo {pages} {503002}
  (\bibinfo {year} {2019})}\BibitemShut {NoStop}%
\bibitem [{\citenamefont {Sadhukhan}\ \emph
  {et~al.}(2024{\natexlab{a}})\citenamefont {Sadhukhan}, \citenamefont {Dey},
  \citenamefont {Karmakar},\ and\ \citenamefont {Nandi}}]{activematterreview}%
  \BibitemOpen
  \bibfield  {author} {\bibinfo {author} {\bibfnamefont {S.}~\bibnamefont
  {Sadhukhan}}, \bibinfo {author} {\bibfnamefont {S.}~\bibnamefont {Dey}},
  \bibinfo {author} {\bibfnamefont {S.}~\bibnamefont {Karmakar}}, \ and\
  \bibinfo {author} {\bibfnamefont {S.~K.}\ \bibnamefont {Nandi}},\ }\href
  {\doibase https://doi.org/10.1140/epjs/s11734-024-01188-1} {\bibfield
  {journal} {\bibinfo  {journal} {The European Physical Journal Special
  Topics}\ ,\ \bibinfo {pages} {1}} (\bibinfo {year}
  {2024}{\natexlab{a}})}\BibitemShut {NoStop}%
\bibitem [{\citenamefont {Poujade}\ \emph {et~al.}(2007)\citenamefont
  {Poujade}, \citenamefont {Grasland-Mongrain}, \citenamefont {Hertzog},
  \citenamefont {Jouanneau}, \citenamefont {Chavrier}, \citenamefont {Ladoux},
  \citenamefont {Buguin},\ and\ \citenamefont {Silberzan}}]{poujade2007}%
  \BibitemOpen
  \bibfield  {author} {\bibinfo {author} {\bibfnamefont {M.}~\bibnamefont
  {Poujade}}, \bibinfo {author} {\bibfnamefont {E.}~\bibnamefont
  {Grasland-Mongrain}}, \bibinfo {author} {\bibfnamefont {A.}~\bibnamefont
  {Hertzog}}, \bibinfo {author} {\bibfnamefont {J.}~\bibnamefont {Jouanneau}},
  \bibinfo {author} {\bibfnamefont {P.}~\bibnamefont {Chavrier}}, \bibinfo
  {author} {\bibfnamefont {B.}~\bibnamefont {Ladoux}}, \bibinfo {author}
  {\bibfnamefont {A.}~\bibnamefont {Buguin}}, \ and\ \bibinfo {author}
  {\bibfnamefont {P.}~\bibnamefont {Silberzan}},\ }\href {\doibase
  10.1073/pnas.0705062104} {\bibfield  {journal} {\bibinfo  {journal} {Proc.
  Natl. Acad. Sci. (USA)}\ }\textbf {\bibinfo {volume} {104}},\ \bibinfo
  {pages} {15988} (\bibinfo {year} {2007})}\BibitemShut {NoStop}%
\bibitem [{\citenamefont {Das}\ \emph {et~al.}(2015)\citenamefont {Das},
  \citenamefont {Safferling}, \citenamefont {Rausch}, \citenamefont {Grabe},
  \citenamefont {Boehm},\ and\ \citenamefont {Spatz}}]{das2015}%
  \BibitemOpen
  \bibfield  {author} {\bibinfo {author} {\bibfnamefont {T.}~\bibnamefont
  {Das}}, \bibinfo {author} {\bibfnamefont {K.}~\bibnamefont {Safferling}},
  \bibinfo {author} {\bibfnamefont {S.}~\bibnamefont {Rausch}}, \bibinfo
  {author} {\bibfnamefont {N.}~\bibnamefont {Grabe}}, \bibinfo {author}
  {\bibfnamefont {H.}~\bibnamefont {Boehm}}, \ and\ \bibinfo {author}
  {\bibfnamefont {J.~P.}\ \bibnamefont {Spatz}},\ }\href {\doibase
  10.1038/ncb3115} {\bibfield  {journal} {\bibinfo  {journal} {Nat. Cell
  Biol.}\ }\textbf {\bibinfo {volume} {17}},\ \bibinfo {pages} {276} (\bibinfo
  {year} {2015})}\BibitemShut {NoStop}%
\bibitem [{\citenamefont {Friedl}\ and\ \citenamefont
  {Gilmour}(2009)}]{friedl2009}%
  \BibitemOpen
  \bibfield  {author} {\bibinfo {author} {\bibfnamefont {P.}~\bibnamefont
  {Friedl}}\ and\ \bibinfo {author} {\bibfnamefont {D.}~\bibnamefont
  {Gilmour}},\ }\href {\doibase https://doi.org/10.1038/nrm2720} {\bibfield
  {journal} {\bibinfo  {journal} {Nature reviews Molecular cell biology}\
  }\textbf {\bibinfo {volume} {10}},\ \bibinfo {pages} {445} (\bibinfo {year}
  {2009})}\BibitemShut {NoStop}%
\bibitem [{\citenamefont {Tambe}\ \emph {et~al.}(2011)\citenamefont {Tambe},
  \citenamefont {Hardin}, \citenamefont {Angelini}, \citenamefont {Rajendran},
  \citenamefont {Park}, \citenamefont {Serra-Picamal}, \citenamefont {Zhou},
  \citenamefont {Zaman}, \citenamefont {Butler}, \citenamefont {Weitz},
  \citenamefont {Fredberg},\ and\ \citenamefont {Trepat}}]{tambe2011}%
  \BibitemOpen
  \bibfield  {author} {\bibinfo {author} {\bibfnamefont {D.~T.}\ \bibnamefont
  {Tambe}}, \bibinfo {author} {\bibfnamefont {C.~C.}\ \bibnamefont {Hardin}},
  \bibinfo {author} {\bibfnamefont {T.~E.}\ \bibnamefont {Angelini}}, \bibinfo
  {author} {\bibfnamefont {K.}~\bibnamefont {Rajendran}}, \bibinfo {author}
  {\bibfnamefont {C.~Y.}\ \bibnamefont {Park}}, \bibinfo {author}
  {\bibfnamefont {X.}~\bibnamefont {Serra-Picamal}}, \bibinfo {author}
  {\bibfnamefont {E.~H.}\ \bibnamefont {Zhou}}, \bibinfo {author}
  {\bibfnamefont {M.~H.}\ \bibnamefont {Zaman}}, \bibinfo {author}
  {\bibfnamefont {J.~P.}\ \bibnamefont {Butler}}, \bibinfo {author}
  {\bibfnamefont {D.~A.}\ \bibnamefont {Weitz}}, \bibinfo {author}
  {\bibfnamefont {J.~J.}\ \bibnamefont {Fredberg}}, \ and\ \bibinfo {author}
  {\bibfnamefont {X.}~\bibnamefont {Trepat}},\ }\href {\doibase
  10.1038/nmat3025} {\bibfield  {journal} {\bibinfo  {journal} {Nat. Mater.}\
  }\textbf {\bibinfo {volume} {10}},\ \bibinfo {pages} {469} (\bibinfo {year}
  {2011})}\BibitemShut {NoStop}%
\bibitem [{\citenamefont {Mongera}\ \emph {et~al.}(2018)\citenamefont
  {Mongera}, \citenamefont {Rowghanian}, \citenamefont {Gustafson},
  \citenamefont {Shelton}, \citenamefont {Kealhofer}, \citenamefont {Carn},
  \citenamefont {Serwane}, \citenamefont {Lucio}, \citenamefont {Giammona},\
  and\ \citenamefont {Camp{\`a}s}}]{mongera2018}%
  \BibitemOpen
  \bibfield  {author} {\bibinfo {author} {\bibfnamefont {A.}~\bibnamefont
  {Mongera}}, \bibinfo {author} {\bibfnamefont {P.}~\bibnamefont {Rowghanian}},
  \bibinfo {author} {\bibfnamefont {H.~J.}\ \bibnamefont {Gustafson}}, \bibinfo
  {author} {\bibfnamefont {E.}~\bibnamefont {Shelton}}, \bibinfo {author}
  {\bibfnamefont {D.~A.}\ \bibnamefont {Kealhofer}}, \bibinfo {author}
  {\bibfnamefont {E.~K.}\ \bibnamefont {Carn}}, \bibinfo {author}
  {\bibfnamefont {F.}~\bibnamefont {Serwane}}, \bibinfo {author} {\bibfnamefont
  {A.~A.}\ \bibnamefont {Lucio}}, \bibinfo {author} {\bibfnamefont
  {J.}~\bibnamefont {Giammona}}, \ and\ \bibinfo {author} {\bibfnamefont
  {O.}~\bibnamefont {Camp{\`a}s}},\ }\href {\doibase 10.1038/s41586-018-0479-2}
  {\bibfield  {journal} {\bibinfo  {journal} {Nature}\ }\textbf {\bibinfo
  {volume} {561}},\ \bibinfo {pages} {401} (\bibinfo {year}
  {2018})}\BibitemShut {NoStop}%
\bibitem [{\citenamefont {Wirtz}\ \emph {et~al.}(2011)\citenamefont {Wirtz},
  \citenamefont {Konstantopoulos},\ and\ \citenamefont {Searson}}]{wirtz2011}%
  \BibitemOpen
  \bibfield  {author} {\bibinfo {author} {\bibfnamefont {D.}~\bibnamefont
  {Wirtz}}, \bibinfo {author} {\bibfnamefont {K.}~\bibnamefont
  {Konstantopoulos}}, \ and\ \bibinfo {author} {\bibfnamefont {P.~C.}\
  \bibnamefont {Searson}},\ }\href {\doibase 10.1038/nrc3080} {\bibfield
  {journal} {\bibinfo  {journal} {Nat. Rev. Canc.}\ }\textbf {\bibinfo {volume}
  {11}},\ \bibinfo {pages} {512} (\bibinfo {year} {2011})}\BibitemShut
  {NoStop}%
\bibitem [{\citenamefont {Park}\ \emph {et~al.}(2016)\citenamefont {Park},
  \citenamefont {Atia}, \citenamefont {Mitchel}, \citenamefont {Fredberg},\
  and\ \citenamefont {Butler}}]{park2016}%
  \BibitemOpen
  \bibfield  {author} {\bibinfo {author} {\bibfnamefont {J.-A.}\ \bibnamefont
  {Park}}, \bibinfo {author} {\bibfnamefont {L.}~\bibnamefont {Atia}}, \bibinfo
  {author} {\bibfnamefont {J.~A.}\ \bibnamefont {Mitchel}}, \bibinfo {author}
  {\bibfnamefont {J.~J.}\ \bibnamefont {Fredberg}}, \ and\ \bibinfo {author}
  {\bibfnamefont {J.~P.}\ \bibnamefont {Butler}},\ }\href {\doibase
  10.1242/jcs.187922} {\bibfield  {journal} {\bibinfo  {journal} {J. Cell
  Sci.}\ }\textbf {\bibinfo {volume} {129}},\ \bibinfo {pages} {3375} (\bibinfo
  {year} {2016})}\BibitemShut {NoStop}%
\bibitem [{\citenamefont {Mitchel}\ \emph {et~al.}(2020)\citenamefont
  {Mitchel}, \citenamefont {Das}, \citenamefont {O’Sullivan}, \citenamefont
  {Stancil}, \citenamefont {DeCamp}, \citenamefont {Koehler}, \citenamefont
  {Ocaña}, \citenamefont {Butler}, \citenamefont {Fredberg}, \citenamefont
  {Nieto}, \citenamefont {Bi},\ and\ \citenamefont {Park}}]{mitchel2020}%
  \BibitemOpen
  \bibfield  {author} {\bibinfo {author} {\bibfnamefont {J.~A.}\ \bibnamefont
  {Mitchel}}, \bibinfo {author} {\bibfnamefont {A.}~\bibnamefont {Das}},
  \bibinfo {author} {\bibfnamefont {M.~J.}\ \bibnamefont {O’Sullivan}},
  \bibinfo {author} {\bibfnamefont {I.~T.}\ \bibnamefont {Stancil}}, \bibinfo
  {author} {\bibfnamefont {S.~J.}\ \bibnamefont {DeCamp}}, \bibinfo {author}
  {\bibfnamefont {S.}~\bibnamefont {Koehler}}, \bibinfo {author} {\bibfnamefont
  {O.~H.}\ \bibnamefont {Ocaña}}, \bibinfo {author} {\bibfnamefont {J.~P.}\
  \bibnamefont {Butler}}, \bibinfo {author} {\bibfnamefont {J.~J.}\
  \bibnamefont {Fredberg}}, \bibinfo {author} {\bibfnamefont {M.~A.}\
  \bibnamefont {Nieto}}, \bibinfo {author} {\bibfnamefont {D.}~\bibnamefont
  {Bi}}, \ and\ \bibinfo {author} {\bibfnamefont {J.-A.}\ \bibnamefont
  {Park}},\ }\href {\doibase 10.1038/s41467-020-18841-7} {\bibfield  {journal}
  {\bibinfo  {journal} {Nat. Commun.}\ }\textbf {\bibinfo {volume} {11}},\
  \bibinfo {pages} {5053} (\bibinfo {year} {2020})}\BibitemShut {NoStop}%
\bibitem [{\citenamefont {Streitberger}\ \emph {et~al.}(2020)\citenamefont
  {Streitberger}, \citenamefont {Lilaj}, \citenamefont {Schrank}, \citenamefont
  {J{\"{u}}rgen~Braun}, \citenamefont {Reiss-Zimmermann}, \citenamefont
  {K{\"{a}}s},\ and\ \citenamefont {Sack}}]{streitberger2020}%
  \BibitemOpen
  \bibfield  {author} {\bibinfo {author} {\bibfnamefont {K.-J.}\ \bibnamefont
  {Streitberger}}, \bibinfo {author} {\bibfnamefont {L.}~\bibnamefont {Lilaj}},
  \bibinfo {author} {\bibfnamefont {F.}~\bibnamefont {Schrank}}, \bibinfo
  {author} {\bibfnamefont {a.~K.-T.~H.}\ \bibnamefont {J{\"{u}}rgen~Braun}},
  \bibinfo {author} {\bibfnamefont {M.}~\bibnamefont {Reiss-Zimmermann}},
  \bibinfo {author} {\bibfnamefont {J.~A.}\ \bibnamefont {K{\"{a}}s}}, \ and\
  \bibinfo {author} {\bibfnamefont {I.}~\bibnamefont {Sack}},\ }\href {\doibase
  10.1073/pnas.1913511116} {\bibfield  {journal} {\bibinfo  {journal} {Proc.
  Natl. Acad. Sci. (USA)}\ }\textbf {\bibinfo {volume} {117}},\ \bibinfo
  {pages} {128} (\bibinfo {year} {2020})}\BibitemShut {NoStop}%
\bibitem [{\citenamefont {Berthier}\ and\ \citenamefont
  {Biroli}(2011)}]{berthier2011}%
  \BibitemOpen
  \bibfield  {author} {\bibinfo {author} {\bibfnamefont {L.}~\bibnamefont
  {Berthier}}\ and\ \bibinfo {author} {\bibfnamefont {G.}~\bibnamefont
  {Biroli}},\ }\href {\doibase 10.1103/RevModPhys.83.587} {\bibfield  {journal}
  {\bibinfo  {journal} {Rev. Mod. Phys.}\ }\textbf {\bibinfo {volume} {83}},\
  \bibinfo {pages} {587} (\bibinfo {year} {2011})}\BibitemShut {NoStop}%
\bibitem [{\citenamefont {Debenedetti}\ and\ \citenamefont
  {Stillinger}(2001)}]{debenedetti2001}%
  \BibitemOpen
  \bibfield  {author} {\bibinfo {author} {\bibfnamefont {P.~G.}\ \bibnamefont
  {Debenedetti}}\ and\ \bibinfo {author} {\bibfnamefont {F.~H.}\ \bibnamefont
  {Stillinger}},\ }\href {\doibase https://doi.org/10.1038/35065704} {\bibfield
   {journal} {\bibinfo  {journal} {Nature}\ }\textbf {\bibinfo {volume}
  {410}},\ \bibinfo {pages} {259} (\bibinfo {year} {2001})}\BibitemShut
  {NoStop}%
\bibitem [{\citenamefont {Trepat}\ \emph {et~al.}(2009)\citenamefont {Trepat},
  \citenamefont {Wasserman}, \citenamefont {Angelini}, \citenamefont {Millet},
  \citenamefont {Weitz}, \citenamefont {Butler},\ and\ \citenamefont
  {Fredberg}}]{trepat2009}%
  \BibitemOpen
  \bibfield  {author} {\bibinfo {author} {\bibfnamefont {X.}~\bibnamefont
  {Trepat}}, \bibinfo {author} {\bibfnamefont {M.~R.}\ \bibnamefont
  {Wasserman}}, \bibinfo {author} {\bibfnamefont {T.~E.}\ \bibnamefont
  {Angelini}}, \bibinfo {author} {\bibfnamefont {E.}~\bibnamefont {Millet}},
  \bibinfo {author} {\bibfnamefont {D.~A.}\ \bibnamefont {Weitz}}, \bibinfo
  {author} {\bibfnamefont {J.~P.}\ \bibnamefont {Butler}}, \ and\ \bibinfo
  {author} {\bibfnamefont {J.~J.}\ \bibnamefont {Fredberg}},\ }\href {\doibase
  10.1038/nphys1269} {\bibfield  {journal} {\bibinfo  {journal} {Nat. Phys.}\
  }\textbf {\bibinfo {volume} {5}},\ \bibinfo {pages} {426} (\bibinfo {year}
  {2009})}\BibitemShut {NoStop}%
\bibitem [{\citenamefont {Zhou}\ \emph {et~al.}(2009)\citenamefont {Zhou},
  \citenamefont {Trepat}, \citenamefont {Park}, \citenamefont {Lenormand},
  \citenamefont {Oliver}, \citenamefont {Mijailovich}, \citenamefont {Hardin},
  \citenamefont {Weitz}, \citenamefont {Butler},\ and\ \citenamefont
  {Fredberg}}]{zhou2009}%
  \BibitemOpen
  \bibfield  {author} {\bibinfo {author} {\bibfnamefont {E.~H.}\ \bibnamefont
  {Zhou}}, \bibinfo {author} {\bibfnamefont {X.}~\bibnamefont {Trepat}},
  \bibinfo {author} {\bibfnamefont {C.~Y.}\ \bibnamefont {Park}}, \bibinfo
  {author} {\bibfnamefont {G.}~\bibnamefont {Lenormand}}, \bibinfo {author}
  {\bibfnamefont {M.~N.}\ \bibnamefont {Oliver}}, \bibinfo {author}
  {\bibfnamefont {S.~M.}\ \bibnamefont {Mijailovich}}, \bibinfo {author}
  {\bibfnamefont {C.}~\bibnamefont {Hardin}}, \bibinfo {author} {\bibfnamefont
  {D.~A.}\ \bibnamefont {Weitz}}, \bibinfo {author} {\bibfnamefont {J.~P.}\
  \bibnamefont {Butler}}, \ and\ \bibinfo {author} {\bibfnamefont {J.~J.}\
  \bibnamefont {Fredberg}},\ }\href {\doibase 10.1073pnas.0901462106}
  {\bibfield  {journal} {\bibinfo  {journal} {Proc. Natl. Acad. Sci. (USA)}\
  }\textbf {\bibinfo {volume} {106}},\ \bibinfo {pages} {10632} (\bibinfo
  {year} {2009})}\BibitemShut {NoStop}%
\bibitem [{\citenamefont {Marchetti}\ \emph {et~al.}(2013)\citenamefont
  {Marchetti}, \citenamefont {Joanny}, \citenamefont {Ramaswamy}, \citenamefont
  {Liverpool}, \citenamefont {Prost}, \citenamefont {Rao},\ and\ \citenamefont
  {Simha}}]{marchetti2013}%
  \BibitemOpen
  \bibfield  {author} {\bibinfo {author} {\bibfnamefont {M.~C.}\ \bibnamefont
  {Marchetti}}, \bibinfo {author} {\bibfnamefont {J.-F.}\ \bibnamefont
  {Joanny}}, \bibinfo {author} {\bibfnamefont {S.}~\bibnamefont {Ramaswamy}},
  \bibinfo {author} {\bibfnamefont {T.~B.}\ \bibnamefont {Liverpool}}, \bibinfo
  {author} {\bibfnamefont {J.}~\bibnamefont {Prost}}, \bibinfo {author}
  {\bibfnamefont {M.}~\bibnamefont {Rao}}, \ and\ \bibinfo {author}
  {\bibfnamefont {R.~A.}\ \bibnamefont {Simha}},\ }\href {\doibase
  https://doi.org/10.1103/RevModPhys.85.1143} {\bibfield  {journal} {\bibinfo
  {journal} {Reviews of modern physics}\ }\textbf {\bibinfo {volume} {85}},\
  \bibinfo {pages} {1143} (\bibinfo {year} {2013})}\BibitemShut {NoStop}%
\bibitem [{\citenamefont {Ramaswamy}(2010)}]{ramaswamy2010}%
  \BibitemOpen
  \bibfield  {author} {\bibinfo {author} {\bibfnamefont {S.}~\bibnamefont
  {Ramaswamy}},\ }\href {\doibase
  https://doi.org/10.1146/annurev-conmatphys-070909-104101} {\bibfield
  {journal} {\bibinfo  {journal} {Annu. Rev. Condens. Matter Phys.}\ }\textbf
  {\bibinfo {volume} {1}},\ \bibinfo {pages} {323} (\bibinfo {year}
  {2010})}\BibitemShut {NoStop}%
\bibitem [{\citenamefont {Bechinger}\ \emph {et~al.}(2016)\citenamefont
  {Bechinger}, \citenamefont {Di~Leonardo}, \citenamefont {L{\"o}wen},
  \citenamefont {Reichhardt}, \citenamefont {Volpe},\ and\ \citenamefont
  {Volpe}}]{bechinger2016}%
  \BibitemOpen
  \bibfield  {author} {\bibinfo {author} {\bibfnamefont {C.}~\bibnamefont
  {Bechinger}}, \bibinfo {author} {\bibfnamefont {R.}~\bibnamefont
  {Di~Leonardo}}, \bibinfo {author} {\bibfnamefont {H.}~\bibnamefont
  {L{\"o}wen}}, \bibinfo {author} {\bibfnamefont {C.}~\bibnamefont
  {Reichhardt}}, \bibinfo {author} {\bibfnamefont {G.}~\bibnamefont {Volpe}}, \
  and\ \bibinfo {author} {\bibfnamefont {G.}~\bibnamefont {Volpe}},\ }\href
  {\doibase https://doi.org/10.1103/RevModPhys.88.045006} {\bibfield  {journal}
  {\bibinfo  {journal} {Reviews of modern physics}\ }\textbf {\bibinfo {volume}
  {88}},\ \bibinfo {pages} {045006} (\bibinfo {year} {2016})}\BibitemShut
  {NoStop}%
\bibitem [{\citenamefont {Vicsek}\ and\ \citenamefont
  {Zafeiris}(2012)}]{vicsek2012}%
  \BibitemOpen
  \bibfield  {author} {\bibinfo {author} {\bibfnamefont {T.}~\bibnamefont
  {Vicsek}}\ and\ \bibinfo {author} {\bibfnamefont {A.}~\bibnamefont
  {Zafeiris}},\ }\href {\doibase https://doi.org/10.1016/j.physrep.2012.03.004}
  {\bibfield  {journal} {\bibinfo  {journal} {Physics reports}\ }\textbf
  {\bibinfo {volume} {517}},\ \bibinfo {pages} {71} (\bibinfo {year}
  {2012})}\BibitemShut {NoStop}%
\bibitem [{\citenamefont {Garcia}\ \emph {et~al.}(2015)\citenamefont {Garcia},
  \citenamefont {Hannezo}, \citenamefont {Elgeti}, \citenamefont {Joanny},
  \citenamefont {Silberzan},\ and\ \citenamefont {Gov}}]{garcia2015}%
  \BibitemOpen
  \bibfield  {author} {\bibinfo {author} {\bibfnamefont {S.}~\bibnamefont
  {Garcia}}, \bibinfo {author} {\bibfnamefont {E.}~\bibnamefont {Hannezo}},
  \bibinfo {author} {\bibfnamefont {J.}~\bibnamefont {Elgeti}}, \bibinfo
  {author} {\bibfnamefont {J.~F.}\ \bibnamefont {Joanny}}, \bibinfo {author}
  {\bibfnamefont {P.}~\bibnamefont {Silberzan}}, \ and\ \bibinfo {author}
  {\bibfnamefont {N.~S.}\ \bibnamefont {Gov}},\ }\href {\doibase
  10.1073/pnas.1510973112} {\bibfield  {journal} {\bibinfo  {journal} {Proc.
  Natl. Acad. Sci. (USA)}\ }\textbf {\bibinfo {volume} {112}},\ \bibinfo
  {pages} {15314} (\bibinfo {year} {2015})}\BibitemShut {NoStop}%
\bibitem [{\citenamefont {Nishizawa}\ \emph {et~al.}(2017)\citenamefont
  {Nishizawa}, \citenamefont {Fujiwara}, \citenamefont {Ikenaga}, \citenamefont
  {Nakajo}, \citenamefont {Yanagisawa},\ and\ \citenamefont
  {Mizuno}}]{nishizawa2017}%
  \BibitemOpen
  \bibfield  {author} {\bibinfo {author} {\bibfnamefont {K.}~\bibnamefont
  {Nishizawa}}, \bibinfo {author} {\bibfnamefont {K.}~\bibnamefont {Fujiwara}},
  \bibinfo {author} {\bibfnamefont {M.}~\bibnamefont {Ikenaga}}, \bibinfo
  {author} {\bibfnamefont {N.}~\bibnamefont {Nakajo}}, \bibinfo {author}
  {\bibfnamefont {M.}~\bibnamefont {Yanagisawa}}, \ and\ \bibinfo {author}
  {\bibfnamefont {D.}~\bibnamefont {Mizuno}},\ }\href {\doibase
  https://doi.org/10.1038/s41598-017-14883-y} {\bibfield  {journal} {\bibinfo
  {journal} {Sci. rep.}\ }\textbf {\bibinfo {volume} {7}},\ \bibinfo {pages}
  {15143} (\bibinfo {year} {2017})}\BibitemShut {NoStop}%
\bibitem [{\citenamefont {Mandal}\ \emph {et~al.}(2016)\citenamefont {Mandal},
  \citenamefont {Bhuyan}, \citenamefont {Rao},\ and\ \citenamefont
  {Dasgupta}}]{mandal2016}%
  \BibitemOpen
  \bibfield  {author} {\bibinfo {author} {\bibfnamefont {R.}~\bibnamefont
  {Mandal}}, \bibinfo {author} {\bibfnamefont {P.~J.}\ \bibnamefont {Bhuyan}},
  \bibinfo {author} {\bibfnamefont {M.}~\bibnamefont {Rao}}, \ and\ \bibinfo
  {author} {\bibfnamefont {C.}~\bibnamefont {Dasgupta}},\ }\href {\doibase
  10.1039/C5SM02950C} {\bibfield  {journal} {\bibinfo  {journal} {Soft Matter}\
  }\textbf {\bibinfo {volume} {12}},\ \bibinfo {pages} {6268} (\bibinfo {year}
  {2016})}\BibitemShut {NoStop}%
\bibitem [{\citenamefont {Flenner}\ \emph {et~al.}(2016)\citenamefont
  {Flenner}, \citenamefont {Szamel},\ and\ \citenamefont
  {Berthier}}]{flenner2016}%
  \BibitemOpen
  \bibfield  {author} {\bibinfo {author} {\bibfnamefont {E.}~\bibnamefont
  {Flenner}}, \bibinfo {author} {\bibfnamefont {G.}~\bibnamefont {Szamel}}, \
  and\ \bibinfo {author} {\bibfnamefont {L.}~\bibnamefont {Berthier}},\ }\href
  {\doibase 10.1039/c6sm01322h} {\bibfield  {journal} {\bibinfo  {journal}
  {Soft Matter}\ }\textbf {\bibinfo {volume} {12}},\ \bibinfo {pages} {7136}
  (\bibinfo {year} {2016})}\BibitemShut {NoStop}%
\bibitem [{\citenamefont {Berthier}\ and\ \citenamefont
  {Kurchan}(2013)}]{berthier2013}%
  \BibitemOpen
  \bibfield  {author} {\bibinfo {author} {\bibfnamefont {L.}~\bibnamefont
  {Berthier}}\ and\ \bibinfo {author} {\bibfnamefont {J.}~\bibnamefont
  {Kurchan}},\ }\href {\doibase 10.1038/nphys2592} {\bibfield  {journal}
  {\bibinfo  {journal} {Nat. Phys.}\ }\textbf {\bibinfo {volume} {9}},\
  \bibinfo {pages} {310} (\bibinfo {year} {2013})}\BibitemShut {NoStop}%
\bibitem [{\citenamefont {Szamel}(2016)}]{szamel2016}%
  \BibitemOpen
  \bibfield  {author} {\bibinfo {author} {\bibfnamefont {G.}~\bibnamefont
  {Szamel}},\ }\href {\doibase 10.1103/PhysRevE.93.012603} {\bibfield
  {journal} {\bibinfo  {journal} {Phys. Rev. E}\ }\textbf {\bibinfo {volume}
  {93}},\ \bibinfo {pages} {012603} (\bibinfo {year} {2016})}\BibitemShut
  {NoStop}%
\bibitem [{\citenamefont {Nandi}\ and\ \citenamefont {Gov}(2017)}]{nandi2017}%
  \BibitemOpen
  \bibfield  {author} {\bibinfo {author} {\bibfnamefont {S.~K.}\ \bibnamefont
  {Nandi}}\ and\ \bibinfo {author} {\bibfnamefont {N.~S.}\ \bibnamefont
  {Gov}},\ }\href {\doibase https://doi.org/10.1039/C7SM01648D} {\bibfield
  {journal} {\bibinfo  {journal} {Soft matter}\ }\textbf {\bibinfo {volume}
  {13}},\ \bibinfo {pages} {7609} (\bibinfo {year} {2017})}\BibitemShut
  {NoStop}%
\bibitem [{\citenamefont {Nandi}\ \emph {et~al.}(2018)\citenamefont {Nandi},
  \citenamefont {Mandal}, \citenamefont {Bhuyan}, \citenamefont {Dasgupta},
  \citenamefont {Rao},\ and\ \citenamefont {Gov}}]{nandi2018}%
  \BibitemOpen
  \bibfield  {author} {\bibinfo {author} {\bibfnamefont {S.~K.}\ \bibnamefont
  {Nandi}}, \bibinfo {author} {\bibfnamefont {R.}~\bibnamefont {Mandal}},
  \bibinfo {author} {\bibfnamefont {P.~J.}\ \bibnamefont {Bhuyan}}, \bibinfo
  {author} {\bibfnamefont {C.}~\bibnamefont {Dasgupta}}, \bibinfo {author}
  {\bibfnamefont {M.}~\bibnamefont {Rao}}, \ and\ \bibinfo {author}
  {\bibfnamefont {N.~S.}\ \bibnamefont {Gov}},\ }\href {\doibase
  10.1073/pnas.1721324115} {\bibfield  {journal} {\bibinfo  {journal}
  {Proceedings of the National Academy of Sciences}\ }\textbf {\bibinfo
  {volume} {115}},\ \bibinfo {pages} {7688} (\bibinfo {year}
  {2018})}\BibitemShut {NoStop}%
\bibitem [{\citenamefont {Paul}\ \emph {et~al.}(2023)\citenamefont {Paul},
  \citenamefont {Mutneja}, \citenamefont {Nandi},\ and\ \citenamefont
  {Karmakar}}]{paul2023}%
  \BibitemOpen
  \bibfield  {author} {\bibinfo {author} {\bibfnamefont {K.}~\bibnamefont
  {Paul}}, \bibinfo {author} {\bibfnamefont {A.}~\bibnamefont {Mutneja}},
  \bibinfo {author} {\bibfnamefont {S.~K.}\ \bibnamefont {Nandi}}, \ and\
  \bibinfo {author} {\bibfnamefont {S.}~\bibnamefont {Karmakar}},\ }\href
  {\doibase 10.1073/pnas.2217073120} {\bibfield  {journal} {\bibinfo  {journal}
  {Proc. Natl. Acad. Sci. (USA)}\ }\textbf {\bibinfo {volume} {120}},\ \bibinfo
  {pages} {e2217073120} (\bibinfo {year} {2023})}\BibitemShut {NoStop}%
\bibitem [{\citenamefont {Keta}\ \emph {et~al.}(2023)\citenamefont {Keta},
  \citenamefont {Mandal}, \citenamefont {Sollich}, \citenamefont {Jack},\ and\
  \citenamefont {Berthier}}]{keta2023}%
  \BibitemOpen
  \bibfield  {author} {\bibinfo {author} {\bibfnamefont {Y.~E.}\ \bibnamefont
  {Keta}}, \bibinfo {author} {\bibfnamefont {R.}~\bibnamefont {Mandal}},
  \bibinfo {author} {\bibfnamefont {P.}~\bibnamefont {Sollich}}, \bibinfo
  {author} {\bibfnamefont {R.~L.}\ \bibnamefont {Jack}}, \ and\ \bibinfo
  {author} {\bibfnamefont {L.}~\bibnamefont {Berthier}},\ }\href {\doibase
  10.1039/d3sm00034f} {\bibfield  {journal} {\bibinfo  {journal} {Soft Matter}\
  }\textbf {\bibinfo {volume} {19}},\ \bibinfo {pages} {3871} (\bibinfo {year}
  {2023})}\BibitemShut {NoStop}%
\bibitem [{\citenamefont {Sadhukhan}\ \emph
  {et~al.}(2024{\natexlab{b}})\citenamefont {Sadhukhan}, \citenamefont {Nandi},
  \citenamefont {Pandey}, \citenamefont {Paoluzzi}, \citenamefont {Dasgupta},
  \citenamefont {Gov},\ and\ \citenamefont {Nandi}}]{sadhukhan2024}%
  \BibitemOpen
  \bibfield  {author} {\bibinfo {author} {\bibfnamefont {S.}~\bibnamefont
  {Sadhukhan}}, \bibinfo {author} {\bibfnamefont {M.~K.}\ \bibnamefont
  {Nandi}}, \bibinfo {author} {\bibfnamefont {S.}~\bibnamefont {Pandey}},
  \bibinfo {author} {\bibfnamefont {M.}~\bibnamefont {Paoluzzi}}, \bibinfo
  {author} {\bibfnamefont {C.}~\bibnamefont {Dasgupta}}, \bibinfo {author}
  {\bibfnamefont {N.~S.}\ \bibnamefont {Gov}}, \ and\ \bibinfo {author}
  {\bibfnamefont {S.~K.}\ \bibnamefont {Nandi}},\ }\href {\doibase
  10.1039/D4SM00352G} {\bibfield  {journal} {\bibinfo  {journal} {Soft Matter}\
  }\textbf {\bibinfo {volume} {20}},\ \bibinfo {pages} {6160} (\bibinfo {year}
  {2024}{\natexlab{b}})}\BibitemShut {NoStop}%
\bibitem [{\citenamefont {Mandal}\ and\ \citenamefont
  {Sollich}(2020)}]{mandal2020PRL}%
  \BibitemOpen
  \bibfield  {author} {\bibinfo {author} {\bibfnamefont {R.}~\bibnamefont
  {Mandal}}\ and\ \bibinfo {author} {\bibfnamefont {P.}~\bibnamefont
  {Sollich}},\ }\href {\doibase 10.1103/PhysRevLett.125.218001} {\bibfield
  {journal} {\bibinfo  {journal} {Phys. Rev. Lett.}\ }\textbf {\bibinfo
  {volume} {125}} (\bibinfo {year} {2020}),\
  10.1103/PhysRevLett.125.218001}\BibitemShut {NoStop}%
\bibitem [{\citenamefont {Mandal}\ and\ \citenamefont
  {Sollich}(2021)}]{mandal2021}%
  \BibitemOpen
  \bibfield  {author} {\bibinfo {author} {\bibfnamefont {R.}~\bibnamefont
  {Mandal}}\ and\ \bibinfo {author} {\bibfnamefont {P.}~\bibnamefont
  {Sollich}},\ }\href {\doibase 10.1088/1361-648X/abef9b} {\bibfield  {journal}
  {\bibinfo  {journal} {J. Phys. Cond. Mat.}\ }\textbf {\bibinfo {volume} {33}}
  (\bibinfo {year} {2021}),\ 10.1088/1361-648X/abef9b}\BibitemShut {NoStop}%
\bibitem [{\citenamefont {Keta}\ \emph {et~al.}(2022)\citenamefont {Keta},
  \citenamefont {Jack},\ and\ \citenamefont {Berthier}}]{keta2022}%
  \BibitemOpen
  \bibfield  {author} {\bibinfo {author} {\bibfnamefont {Y.~E.}\ \bibnamefont
  {Keta}}, \bibinfo {author} {\bibfnamefont {R.~L.}\ \bibnamefont {Jack}}, \
  and\ \bibinfo {author} {\bibfnamefont {L.}~\bibnamefont {Berthier}},\ }\href
  {\doibase 10.1103/PhysRevLett.129.048002} {\bibfield  {journal} {\bibinfo
  {journal} {Phys. Rev. Lett.}\ }\textbf {\bibinfo {volume} {129}} (\bibinfo
  {year} {2022}),\ 10.1103/PhysRevLett.129.048002}\BibitemShut {NoStop}%
\bibitem [{\citenamefont {Zheng}\ \emph {et~al.}(2024)\citenamefont {Zheng},
  \citenamefont {Khomenko},\ and\ \citenamefont {Charbonneau}}]{zheng2024}%
  \BibitemOpen
  \bibfield  {author} {\bibinfo {author} {\bibfnamefont {M.}~\bibnamefont
  {Zheng}}, \bibinfo {author} {\bibfnamefont {D.}~\bibnamefont {Khomenko}}, \
  and\ \bibinfo {author} {\bibfnamefont {P.}~\bibnamefont {Charbonneau}},\
  }\href {http://arxiv.org/abs/2409.12037} {\bibfield  {journal} {\bibinfo
  {journal} {arXiv}\ ,\ \bibinfo {pages} {2409.12037}} (\bibinfo {year}
  {2024})}\BibitemShut {NoStop}%
\bibitem [{\citenamefont {Kolya}\ \emph {et~al.}(2024)\citenamefont {Kolya},
  \citenamefont {Pareek},\ and\ \citenamefont {Nandi}}]{kolya2024}%
  \BibitemOpen
  \bibfield  {author} {\bibinfo {author} {\bibfnamefont {S.}~\bibnamefont
  {Kolya}}, \bibinfo {author} {\bibfnamefont {P.}~\bibnamefont {Pareek}}, \
  and\ \bibinfo {author} {\bibfnamefont {S.~K.}\ \bibnamefont {Nandi}},\ }\href
  {http://arxiv.org/abs/2410.15928} {\bibfield  {journal} {\bibinfo  {journal}
  {arXiv}\ ,\ \bibinfo {pages} {2410.15928}} (\bibinfo {year}
  {2024})}\BibitemShut {NoStop}%
\bibitem [{\citenamefont {Langer}(2014)}]{langer2014}%
  \BibitemOpen
  \bibfield  {author} {\bibinfo {author} {\bibfnamefont {J.~S.}\ \bibnamefont
  {Langer}},\ }\href {\doibase 10.1088/0034-4885/77/4/042501} {\enquote
  {\bibinfo {title} {Theories of glass formation and the glass transition},}\ }
  (\bibinfo {year} {2014})\BibitemShut {NoStop}%
\bibitem [{\citenamefont {Biroli}\ \emph {et~al.}(2022)\citenamefont {Biroli},
  \citenamefont {Charbonneau}, \citenamefont {Folena}, \citenamefont {Hu},\
  and\ \citenamefont {Zamponi}}]{biroli2022}%
  \BibitemOpen
  \bibfield  {author} {\bibinfo {author} {\bibfnamefont {G.}~\bibnamefont
  {Biroli}}, \bibinfo {author} {\bibfnamefont {P.}~\bibnamefont {Charbonneau}},
  \bibinfo {author} {\bibfnamefont {G.}~\bibnamefont {Folena}}, \bibinfo
  {author} {\bibfnamefont {Y.}~\bibnamefont {Hu}}, \ and\ \bibinfo {author}
  {\bibfnamefont {F.}~\bibnamefont {Zamponi}},\ }\href {\doibase
  10.1103/PhysRevLett.128.175501} {\bibfield  {journal} {\bibinfo  {journal}
  {Phys. Rev. Lett.}\ }\textbf {\bibinfo {volume} {128}},\ \bibinfo {pages}
  {175501} (\bibinfo {year} {2022})}\BibitemShut {NoStop}%
\bibitem [{\citenamefont {Lemaître}\ and\ \citenamefont
  {Caroli}(2009)}]{lemaitre2009}%
  \BibitemOpen
  \bibfield  {author} {\bibinfo {author} {\bibfnamefont {A.}~\bibnamefont
  {Lemaître}}\ and\ \bibinfo {author} {\bibfnamefont {C.}~\bibnamefont
  {Caroli}},\ }\href {\doibase 10.1103/PhysRevLett.103.065501} {\bibfield
  {journal} {\bibinfo  {journal} {Phys. Rev. Lett.}\ }\textbf {\bibinfo
  {volume} {103}},\ \bibinfo {pages} {065501} (\bibinfo {year}
  {2009})}\BibitemShut {NoStop}%
\bibitem [{\citenamefont {Lemaître}(2014)}]{lemaitre2014}%
  \BibitemOpen
  \bibfield  {author} {\bibinfo {author} {\bibfnamefont {A.}~\bibnamefont
  {Lemaître}},\ }\href {\doibase 10.1103/PhysRevLett.113.245702} {\bibfield
  {journal} {\bibinfo  {journal} {Phys. Rev. Lett.}\ }\textbf {\bibinfo
  {volume} {113}},\ \bibinfo {pages} {245702} (\bibinfo {year}
  {2014})}\BibitemShut {NoStop}%
\bibitem [{\citenamefont {Martens}\ \emph {et~al.}(2012)\citenamefont
  {Martens}, \citenamefont {Bocquet},\ and\ \citenamefont
  {Barrat}}]{martens2012}%
  \BibitemOpen
  \bibfield  {author} {\bibinfo {author} {\bibfnamefont {K.}~\bibnamefont
  {Martens}}, \bibinfo {author} {\bibfnamefont {L.}~\bibnamefont {Bocquet}}, \
  and\ \bibinfo {author} {\bibfnamefont {J.-L.}\ \bibnamefont {Barrat}},\
  }\href {\doibase https://doi.org/10.1039/C2SM07090A} {\bibfield  {journal}
  {\bibinfo  {journal} {Soft Matter}\ }\textbf {\bibinfo {volume} {8}},\
  \bibinfo {pages} {4197} (\bibinfo {year} {2012})}\BibitemShut {NoStop}%
\bibitem [{\citenamefont {Ferrero}\ \emph {et~al.}(2014)\citenamefont
  {Ferrero}, \citenamefont {Martens},\ and\ \citenamefont
  {Barrat}}]{ferrero2014}%
  \BibitemOpen
  \bibfield  {author} {\bibinfo {author} {\bibfnamefont {E.~E.}\ \bibnamefont
  {Ferrero}}, \bibinfo {author} {\bibfnamefont {K.}~\bibnamefont {Martens}}, \
  and\ \bibinfo {author} {\bibfnamefont {J.-L.}\ \bibnamefont {Barrat}},\
  }\href {\doibase https://doi.org/10.1103/PhysRevLett.113.248301} {\bibfield
  {journal} {\bibinfo  {journal} {Physical review letters}\ }\textbf {\bibinfo
  {volume} {113}},\ \bibinfo {pages} {248301} (\bibinfo {year}
  {2014})}\BibitemShut {NoStop}%
\bibitem [{\citenamefont {Lin}\ \emph {et~al.}(2014)\citenamefont {Lin},
  \citenamefont {Lerner}, \citenamefont {Rosso},\ and\ \citenamefont
  {Wyart}}]{lin2014}%
  \BibitemOpen
  \bibfield  {author} {\bibinfo {author} {\bibfnamefont {J.}~\bibnamefont
  {Lin}}, \bibinfo {author} {\bibfnamefont {E.}~\bibnamefont {Lerner}},
  \bibinfo {author} {\bibfnamefont {A.}~\bibnamefont {Rosso}}, \ and\ \bibinfo
  {author} {\bibfnamefont {M.}~\bibnamefont {Wyart}},\ }\href {\doibase
  https://doi.org/10.1073/pnas.1406391111} {\bibfield  {journal} {\bibinfo
  {journal} {Proceedings of the National Academy of Sciences}\ }\textbf
  {\bibinfo {volume} {111}},\ \bibinfo {pages} {14382} (\bibinfo {year}
  {2014})}\BibitemShut {NoStop}%
\bibitem [{\citenamefont {Nicolas}\ \emph {et~al.}(2018)\citenamefont
  {Nicolas}, \citenamefont {Ferrero}, \citenamefont {Martens},\ and\
  \citenamefont {Barrat}}]{nicolas2018}%
  \BibitemOpen
  \bibfield  {author} {\bibinfo {author} {\bibfnamefont {A.}~\bibnamefont
  {Nicolas}}, \bibinfo {author} {\bibfnamefont {E.~E.}\ \bibnamefont
  {Ferrero}}, \bibinfo {author} {\bibfnamefont {K.}~\bibnamefont {Martens}}, \
  and\ \bibinfo {author} {\bibfnamefont {J.-L.}\ \bibnamefont {Barrat}},\
  }\href {\doibase 10.1103/RevModPhys.90.045006} {\bibfield  {journal}
  {\bibinfo  {journal} {Rev. Mod. Phys.}\ }\textbf {\bibinfo {volume} {90}},\
  \bibinfo {pages} {045006} (\bibinfo {year} {2018})}\BibitemShut {NoStop}%
\bibitem [{\citenamefont {Dyre}(2006)}]{dyre2006}%
  \BibitemOpen
  \bibfield  {author} {\bibinfo {author} {\bibfnamefont {J.~C.}\ \bibnamefont
  {Dyre}},\ }\href {\doibase https://doi.org/10.1103/RevModPhys.78.953}
  {\bibfield  {journal} {\bibinfo  {journal} {Reviews of modern physics}\
  }\textbf {\bibinfo {volume} {78}},\ \bibinfo {pages} {953} (\bibinfo {year}
  {2006})}\BibitemShut {NoStop}%
\bibitem [{\citenamefont {Chattoraj}\ and\ \citenamefont
  {Lemaître}(2013)}]{chattoraj2013}%
  \BibitemOpen
  \bibfield  {author} {\bibinfo {author} {\bibfnamefont {J.}~\bibnamefont
  {Chattoraj}}\ and\ \bibinfo {author} {\bibfnamefont {A.}~\bibnamefont
  {Lemaître}},\ }\href {\doibase 10.1103/PhysRevLett.111.066001} {\bibfield
  {journal} {\bibinfo  {journal} {Phys. Rev. Lett.}\ }\textbf {\bibinfo
  {volume} {111}},\ \bibinfo {pages} {066001} (\bibinfo {year}
  {2013})}\BibitemShut {NoStop}%
\bibitem [{\citenamefont {Dyre}(2024)}]{dyre2024}%
  \BibitemOpen
  \bibfield  {author} {\bibinfo {author} {\bibfnamefont {J.~C.}\ \bibnamefont
  {Dyre}},\ }\href {\doibase https://doi.org/10.1021/acs.jpclett.3c03308}
  {\bibfield  {journal} {\bibinfo  {journal} {The Journal of Physical Chemistry
  Letters}\ }\textbf {\bibinfo {volume} {15}},\ \bibinfo {pages} {1603}
  (\bibinfo {year} {2024})}\BibitemShut {NoStop}%
\bibitem [{\citenamefont {Schr{\o}der}\ and\ \citenamefont
  {Dyre}(2020)}]{schroder2020}%
  \BibitemOpen
  \bibfield  {author} {\bibinfo {author} {\bibfnamefont {T.~B.}\ \bibnamefont
  {Schr{\o}der}}\ and\ \bibinfo {author} {\bibfnamefont {J.~C.}\ \bibnamefont
  {Dyre}},\ }\href {\doibase https://doi.org/10.1063/5.0004093} {\bibfield
  {journal} {\bibinfo  {journal} {The Journal of chemical physics}\ }\textbf
  {\bibinfo {volume} {152}} (\bibinfo {year} {2020}),\
  https://doi.org/10.1063/5.0004093}\BibitemShut {NoStop}%
\bibitem [{\citenamefont {Eshelby}(1957)}]{eshelby1957}%
  \BibitemOpen
  \bibfield  {author} {\bibinfo {author} {\bibfnamefont {J.~D.}\ \bibnamefont
  {Eshelby}},\ }\href {\doibase 10.1098/rspa.1957.0133} {\bibfield  {journal}
  {\bibinfo  {journal} {Proc. Roy. Soc. London A}\ }\textbf {\bibinfo {volume}
  {241}},\ \bibinfo {pages} {376} (\bibinfo {year} {1957})}\BibitemShut
  {NoStop}%
\bibitem [{\citenamefont {Bulatov}\ and\ \citenamefont
  {Argon}(1994)}]{bulatov1994}%
  \BibitemOpen
  \bibfield  {author} {\bibinfo {author} {\bibfnamefont {V.}~\bibnamefont
  {Bulatov}}\ and\ \bibinfo {author} {\bibfnamefont {A.}~\bibnamefont
  {Argon}},\ }\href {\doibase 10.1088/0965-0393/2/2/001} {\bibfield  {journal}
  {\bibinfo  {journal} {Mod. Sim. Mat. Sci. Eng.}\ }\textbf {\bibinfo {volume}
  {2}},\ \bibinfo {pages} {167} (\bibinfo {year} {1994})}\BibitemShut {NoStop}%
\bibitem [{\citenamefont {Ozawa}\ \emph {et~al.}(2018)\citenamefont {Ozawa},
  \citenamefont {Berthier}, \citenamefont {Biroli}, \citenamefont {Rosso},\
  and\ \citenamefont {Tarjus}}]{ozawa2018}%
  \BibitemOpen
  \bibfield  {author} {\bibinfo {author} {\bibfnamefont {M.}~\bibnamefont
  {Ozawa}}, \bibinfo {author} {\bibfnamefont {L.}~\bibnamefont {Berthier}},
  \bibinfo {author} {\bibfnamefont {G.}~\bibnamefont {Biroli}}, \bibinfo
  {author} {\bibfnamefont {A.}~\bibnamefont {Rosso}}, \ and\ \bibinfo {author}
  {\bibfnamefont {G.}~\bibnamefont {Tarjus}},\ }\href {\doibase
  https://doi.org/10.1073/pnas.1806156115} {\bibfield  {journal} {\bibinfo
  {journal} {Proceedings of the National Academy of Sciences}\ }\textbf
  {\bibinfo {volume} {115}},\ \bibinfo {pages} {6656} (\bibinfo {year}
  {2018})}\BibitemShut {NoStop}%
\bibitem [{\citenamefont {Rossi}\ \emph {et~al.}(2022)\citenamefont {Rossi},
  \citenamefont {Biroli}, \citenamefont {Ozawa}, \citenamefont {Tarjus},\ and\
  \citenamefont {Zamponi}}]{rossi2023}%
  \BibitemOpen
  \bibfield  {author} {\bibinfo {author} {\bibfnamefont {S.}~\bibnamefont
  {Rossi}}, \bibinfo {author} {\bibfnamefont {G.}~\bibnamefont {Biroli}},
  \bibinfo {author} {\bibfnamefont {M.}~\bibnamefont {Ozawa}}, \bibinfo
  {author} {\bibfnamefont {G.}~\bibnamefont {Tarjus}}, \ and\ \bibinfo {author}
  {\bibfnamefont {F.}~\bibnamefont {Zamponi}},\ }\href {\doibase
  10.1103/PhysRevLett.129.228002} {\bibfield  {journal} {\bibinfo  {journal}
  {Phys. Rev. Lett.}\ }\textbf {\bibinfo {volume} {129}},\ \bibinfo {pages}
  {228002} (\bibinfo {year} {2022})}\BibitemShut {NoStop}%
\bibitem [{\citenamefont {Picard}\ \emph {et~al.}(2004)\citenamefont {Picard},
  \citenamefont {Ajdari}, \citenamefont {Lequeux},\ and\ \citenamefont
  {Bocquet}}]{picard2004}%
  \BibitemOpen
  \bibfield  {author} {\bibinfo {author} {\bibfnamefont {G.}~\bibnamefont
  {Picard}}, \bibinfo {author} {\bibfnamefont {A.}~\bibnamefont {Ajdari}},
  \bibinfo {author} {\bibfnamefont {F.}~\bibnamefont {Lequeux}}, \ and\
  \bibinfo {author} {\bibfnamefont {L.}~\bibnamefont {Bocquet}},\ }\href
  {\doibase https://doi.org/10.1140/epje/i2004-10054-8} {\bibfield  {journal}
  {\bibinfo  {journal} {The European Physical Journal E}\ }\textbf {\bibinfo
  {volume} {15}},\ \bibinfo {pages} {371} (\bibinfo {year} {2004})}\BibitemShut
  {NoStop}%
\bibitem [{\citenamefont {Ozawa}\ and\ \citenamefont
  {Biroli}(2023)}]{ozawa2023}%
  \BibitemOpen
  \bibfield  {author} {\bibinfo {author} {\bibfnamefont {M.}~\bibnamefont
  {Ozawa}}\ and\ \bibinfo {author} {\bibfnamefont {G.}~\bibnamefont {Biroli}},\
  }\href {\doibase 10.1103/PhysRevLett.130.138201} {\bibfield  {journal}
  {\bibinfo  {journal} {Phys. Rev. Lett.}\ }\textbf {\bibinfo {volume} {130}},\
  \bibinfo {pages} {138201} (\bibinfo {year} {2023})}\BibitemShut {NoStop}%
\bibitem [{\citenamefont {Tahaei}\ \emph {et~al.}(2023)\citenamefont {Tahaei},
  \citenamefont {Biroli}, \citenamefont {Ozawa}, \citenamefont {Popovi{\'c}},\
  and\ \citenamefont {Wyart}}]{tahaei2023}%
  \BibitemOpen
  \bibfield  {author} {\bibinfo {author} {\bibfnamefont {A.}~\bibnamefont
  {Tahaei}}, \bibinfo {author} {\bibfnamefont {G.}~\bibnamefont {Biroli}},
  \bibinfo {author} {\bibfnamefont {M.}~\bibnamefont {Ozawa}}, \bibinfo
  {author} {\bibfnamefont {M.}~\bibnamefont {Popovi{\'c}}}, \ and\ \bibinfo
  {author} {\bibfnamefont {M.}~\bibnamefont {Wyart}},\ }\href {\doibase
  https://doi.org/10.1103/PhysRevX.13.031034} {\bibfield  {journal} {\bibinfo
  {journal} {Physical Review X}\ }\textbf {\bibinfo {volume} {13}},\ \bibinfo
  {pages} {031034} (\bibinfo {year} {2023})}\BibitemShut {NoStop}%
\bibitem [{\citenamefont {Popovi\ifmmode~\acute{c}\else \'{c}\fi{}}\ \emph
  {et~al.}(2021)\citenamefont {Popovi\ifmmode~\acute{c}\else \'{c}\fi{}},
  \citenamefont {de~Geus}, \citenamefont {Ji},\ and\ \citenamefont
  {Wyart}}]{popovic2021}%
  \BibitemOpen
  \bibfield  {author} {\bibinfo {author} {\bibfnamefont {M.}~\bibnamefont
  {Popovi\ifmmode~\acute{c}\else \'{c}\fi{}}}, \bibinfo {author} {\bibfnamefont
  {T.~W.~J.}\ \bibnamefont {de~Geus}}, \bibinfo {author} {\bibfnamefont
  {W.}~\bibnamefont {Ji}}, \ and\ \bibinfo {author} {\bibfnamefont
  {M.}~\bibnamefont {Wyart}},\ }\href {\doibase
  https://doi.org/10.1103/PhysRevE.104.025010} {\bibfield  {journal} {\bibinfo
  {journal} {Physical Review E}\ }\textbf {\bibinfo {volume} {104}} (\bibinfo
  {year} {2021}),\ https://doi.org/10.1103/PhysRevE.104.025010}\BibitemShut
  {NoStop}%
\bibitem [{\citenamefont {Ferrero}\ \emph {et~al.}(2021)\citenamefont
  {Ferrero}, \citenamefont {Kolton},\ and\ \citenamefont
  {Jagla}}]{ferrero2021}%
  \BibitemOpen
  \bibfield  {author} {\bibinfo {author} {\bibfnamefont {E.~E.}\ \bibnamefont
  {Ferrero}}, \bibinfo {author} {\bibfnamefont {A.~B.}\ \bibnamefont {Kolton}},
  \ and\ \bibinfo {author} {\bibfnamefont {E.~A.}\ \bibnamefont {Jagla}},\
  }\href {\doibase https://doi.org/10.1103/PhysRevMaterials.5.115602}
  {\bibfield  {journal} {\bibinfo  {journal} {Physical Review Materials}\
  }\textbf {\bibinfo {volume} {5}},\ \bibinfo {pages} {115602} (\bibinfo {year}
  {2021})}\BibitemShut {NoStop}%
\bibitem [{\citenamefont {Ben-Isaac}\ \emph {et~al.}(2015)\citenamefont
  {Ben-Isaac}, \citenamefont {Fodor}, \citenamefont {Visco}, \citenamefont {van
  Wijland},\ and\ \citenamefont {Gov}}]{benisaac2015}%
  \BibitemOpen
  \bibfield  {author} {\bibinfo {author} {\bibfnamefont {E.}~\bibnamefont
  {Ben-Isaac}}, \bibinfo {author} {\bibfnamefont {{\'{E}}.}~\bibnamefont
  {Fodor}}, \bibinfo {author} {\bibfnamefont {P.}~\bibnamefont {Visco}},
  \bibinfo {author} {\bibfnamefont {F.}~\bibnamefont {van Wijland}}, \ and\
  \bibinfo {author} {\bibfnamefont {N.~S.}\ \bibnamefont {Gov}},\ }\href
  {\doibase 10.1103/PhysRevE.92.012716} {\bibfield  {journal} {\bibinfo
  {journal} {Phys. Rev. E}\ }\textbf {\bibinfo {volume} {92}},\ \bibinfo
  {pages} {012716} (\bibinfo {year} {2015})}\BibitemShut {NoStop}%
\bibitem [{\citenamefont {Woillez}\ \emph {et~al.}(2019)\citenamefont
  {Woillez}, \citenamefont {Zhao}, \citenamefont {Kafri}, \citenamefont
  {Lecomte},\ and\ \citenamefont {Tailleur}}]{woillez2019}%
  \BibitemOpen
  \bibfield  {author} {\bibinfo {author} {\bibfnamefont {E.}~\bibnamefont
  {Woillez}}, \bibinfo {author} {\bibfnamefont {Y.}~\bibnamefont {Zhao}},
  \bibinfo {author} {\bibfnamefont {Y.}~\bibnamefont {Kafri}}, \bibinfo
  {author} {\bibfnamefont {V.}~\bibnamefont {Lecomte}}, \ and\ \bibinfo
  {author} {\bibfnamefont {J.}~\bibnamefont {Tailleur}},\ }\href {\doibase
  10.1103/PhysRevLett.122.258001} {\bibfield  {journal} {\bibinfo  {journal}
  {Phys. Rev. Lett.}\ }\textbf {\bibinfo {volume} {122}},\ \bibinfo {pages}
  {258001} (\bibinfo {year} {2019})}\BibitemShut {NoStop}%
\bibitem [{\citenamefont {Caprini}\ \emph {et~al.}(2019)\citenamefont
  {Caprini}, \citenamefont {Marconi}, \citenamefont {Puglisi},\ and\
  \citenamefont {Vulpiani}}]{caprini2019}%
  \BibitemOpen
  \bibfield  {author} {\bibinfo {author} {\bibfnamefont {L.}~\bibnamefont
  {Caprini}}, \bibinfo {author} {\bibfnamefont {U.~M.~B.}\ \bibnamefont
  {Marconi}}, \bibinfo {author} {\bibfnamefont {A.}~\bibnamefont {Puglisi}}, \
  and\ \bibinfo {author} {\bibfnamefont {A.}~\bibnamefont {Vulpiani}},\ }\href
  {\doibase 10.1063/1.5080537} {\bibfield  {journal} {\bibinfo  {journal} {J.
  Chem. Phys.}\ }\textbf {\bibinfo {volume} {150}},\ \bibinfo {pages} {024902}
  (\bibinfo {year} {2019})}\BibitemShut {NoStop}%
\bibitem [{\citenamefont {Maloney}\ and\ \citenamefont
  {Lacks}(2006)}]{maloney2006}%
  \BibitemOpen
  \bibfield  {author} {\bibinfo {author} {\bibfnamefont {C.~E.}\ \bibnamefont
  {Maloney}}\ and\ \bibinfo {author} {\bibfnamefont {D.~J.}\ \bibnamefont
  {Lacks}},\ }\href {\doibase 10.1103/PhysRevE.73.061106} {\bibfield  {journal}
  {\bibinfo  {journal} {Phys. Rev. E}\ }\textbf {\bibinfo {volume} {73}},\
  \bibinfo {pages} {061106} (\bibinfo {year} {2006})}\BibitemShut {NoStop}%
\bibitem [{\citenamefont {Lerbinger}\ \emph {et~al.}(2022)\citenamefont
  {Lerbinger}, \citenamefont {Barbot}, \citenamefont {Vandembroucq},\ and\
  \citenamefont {Patinet}}]{lerbinger2022}%
  \BibitemOpen
  \bibfield  {author} {\bibinfo {author} {\bibfnamefont {M.}~\bibnamefont
  {Lerbinger}}, \bibinfo {author} {\bibfnamefont {A.}~\bibnamefont {Barbot}},
  \bibinfo {author} {\bibfnamefont {D.}~\bibnamefont {Vandembroucq}}, \ and\
  \bibinfo {author} {\bibfnamefont {S.}~\bibnamefont {Patinet}},\ }\href
  {\doibase 10.1103/PhysRevLett.129.195501} {\bibfield  {journal} {\bibinfo
  {journal} {Phys. Rev. Lett.}\ }\textbf {\bibinfo {volume} {129}},\ \bibinfo
  {pages} {195501} (\bibinfo {year} {2022})}\BibitemShut {NoStop}%
\bibitem [{\citenamefont {Barbot}\ \emph {et~al.}(2018)\citenamefont {Barbot},
  \citenamefont {Lerbinger}, \citenamefont {Hernandez-Garcia}, \citenamefont
  {Garc{\'\i}a-Garc{\'\i}a}, \citenamefont {Falk}, \citenamefont
  {Vandembroucq},\ and\ \citenamefont {Patinet}}]{barbot2018}%
  \BibitemOpen
  \bibfield  {author} {\bibinfo {author} {\bibfnamefont {A.}~\bibnamefont
  {Barbot}}, \bibinfo {author} {\bibfnamefont {M.}~\bibnamefont {Lerbinger}},
  \bibinfo {author} {\bibfnamefont {A.}~\bibnamefont {Hernandez-Garcia}},
  \bibinfo {author} {\bibfnamefont {R.}~\bibnamefont
  {Garc{\'\i}a-Garc{\'\i}a}}, \bibinfo {author} {\bibfnamefont {M.~L.}\
  \bibnamefont {Falk}}, \bibinfo {author} {\bibfnamefont {D.}~\bibnamefont
  {Vandembroucq}}, \ and\ \bibinfo {author} {\bibfnamefont {S.}~\bibnamefont
  {Patinet}},\ }\href@noop {} {\bibfield  {journal} {\bibinfo  {journal}
  {Physical Review E}\ }\textbf {\bibinfo {volume} {97}},\ \bibinfo {pages}
  {033001} (\bibinfo {year} {2018})}\BibitemShut {NoStop}%
\bibitem [{\citenamefont {Berthier}\ \emph {et~al.}(2017)\citenamefont
  {Berthier}, \citenamefont {Flenner},\ and\ \citenamefont
  {Szamel}}]{Berthier2017}%
  \BibitemOpen
  \bibfield  {author} {\bibinfo {author} {\bibfnamefont {L.}~\bibnamefont
  {Berthier}}, \bibinfo {author} {\bibfnamefont {E.}~\bibnamefont {Flenner}}, \
  and\ \bibinfo {author} {\bibfnamefont {G.}~\bibnamefont {Szamel}},\ }\href
  {\doibase 10.1088/1367-2630/aa914e} {\bibfield  {journal} {\bibinfo
  {journal} {New J. Phys.}\ }\textbf {\bibinfo {volume} {19}},\ \bibinfo
  {pages} {125006} (\bibinfo {year} {2017})}\BibitemShut {NoStop}%
\bibitem [{\citenamefont {Paoluzzi}\ \emph {et~al.}(2022)\citenamefont
  {Paoluzzi}, \citenamefont {Levis},\ and\ \citenamefont
  {Pagonabarraga}}]{matteo2022}%
  \BibitemOpen
  \bibfield  {author} {\bibinfo {author} {\bibfnamefont {M.}~\bibnamefont
  {Paoluzzi}}, \bibinfo {author} {\bibfnamefont {D.}~\bibnamefont {Levis}}, \
  and\ \bibinfo {author} {\bibfnamefont {I.}~\bibnamefont {Pagonabarraga}},\
  }\href {\doibase 10.1038/s42005-022-00886-3} {\bibfield  {journal} {\bibinfo
  {journal} {Comm. Phys.}\ }\textbf {\bibinfo {volume} {5}},\ \bibinfo {pages}
  {111} (\bibinfo {year} {2022})}\BibitemShut {NoStop}%
\bibitem [{\citenamefont {Ghoshal}\ and\ \citenamefont
  {Joy}(2020)}]{ghoshal2020}%
  \BibitemOpen
  \bibfield  {author} {\bibinfo {author} {\bibfnamefont {D.}~\bibnamefont
  {Ghoshal}}\ and\ \bibinfo {author} {\bibfnamefont {A.}~\bibnamefont {Joy}},\
  }\href {\doibase 10.1103/physreve.102.062605} {\bibfield  {journal} {\bibinfo
   {journal} {Phys. Rev. E}\ }\textbf {\bibinfo {volume} {102}},\ \bibinfo
  {pages} {062605} (\bibinfo {year} {2020})}\BibitemShut {NoStop}%
\bibitem [{\citenamefont {Dasgupta}\ \emph {et~al.}(1991)\citenamefont
  {Dasgupta}, \citenamefont {Indrani}, \citenamefont {Ramaswamy},\ and\
  \citenamefont {Phani}}]{dasgupta1991}%
  \BibitemOpen
  \bibfield  {author} {\bibinfo {author} {\bibfnamefont {C.}~\bibnamefont
  {Dasgupta}}, \bibinfo {author} {\bibfnamefont {A.~V.}\ \bibnamefont
  {Indrani}}, \bibinfo {author} {\bibfnamefont {S.}~\bibnamefont {Ramaswamy}},
  \ and\ \bibinfo {author} {\bibfnamefont {M.~K.}\ \bibnamefont {Phani}},\
  }\href {\doibase 10.1209/0295-5075/15/3/013} {\bibfield  {journal} {\bibinfo
  {journal} {Europhys. Lett.}\ }\textbf {\bibinfo {volume} {15}},\ \bibinfo
  {pages} {307} (\bibinfo {year} {1991})}\BibitemShut {NoStop}%
\bibitem [{\citenamefont {Biroli}\ \emph {et~al.}(2006)\citenamefont {Biroli},
  \citenamefont {Bouchaud}, \citenamefont {Miyazaki},\ and\ \citenamefont
  {Reichman}}]{IMCT}%
  \BibitemOpen
  \bibfield  {author} {\bibinfo {author} {\bibfnamefont {G.}~\bibnamefont
  {Biroli}}, \bibinfo {author} {\bibfnamefont {J.-P.}\ \bibnamefont
  {Bouchaud}}, \bibinfo {author} {\bibfnamefont {K.}~\bibnamefont {Miyazaki}},
  \ and\ \bibinfo {author} {\bibfnamefont {D.~R.}\ \bibnamefont {Reichman}},\
  }\href {\doibase 10.1103/PhysRevLett.97.195701} {\bibfield  {journal}
  {\bibinfo  {journal} {Phys. Rev. Lett.}\ }\textbf {\bibinfo {volume} {97}},\
  \bibinfo {pages} {195701} (\bibinfo {year} {2006})}\BibitemShut {NoStop}%
\bibitem [{\citenamefont {Mandal}\ \emph {et~al.}(2022)\citenamefont {Mandal},
  \citenamefont {Nandi}, \citenamefont {Dasgupta}, \citenamefont {Sollich},\
  and\ \citenamefont {Gov}}]{mandal2022}%
  \BibitemOpen
  \bibfield  {author} {\bibinfo {author} {\bibfnamefont {R.}~\bibnamefont
  {Mandal}}, \bibinfo {author} {\bibfnamefont {S.~K.}\ \bibnamefont {Nandi}},
  \bibinfo {author} {\bibfnamefont {C.}~\bibnamefont {Dasgupta}}, \bibinfo
  {author} {\bibfnamefont {P.}~\bibnamefont {Sollich}}, \ and\ \bibinfo
  {author} {\bibfnamefont {N.~S.}\ \bibnamefont {Gov}},\ }\href {\doibase
  10.1088/2399-6528/ac9c47} {\bibfield  {journal} {\bibinfo  {journal} {J.
  Phys. Commun.}\ }\textbf {\bibinfo {volume} {6}},\ \bibinfo {pages} {115001}
  (\bibinfo {year} {2022})}\BibitemShut {NoStop}%
\bibitem [{\citenamefont {Angell}(1991)}]{angell1991}%
  \BibitemOpen
  \bibfield  {author} {\bibinfo {author} {\bibfnamefont {C.}~\bibnamefont
  {Angell}},\ }\href {\doibase https://doi.org/10.1016/0022-3093(91)90266-9}
  {\bibfield  {journal} {\bibinfo  {journal} {J. Non-Crys. Solids}\ }\textbf
  {\bibinfo {volume} {131-133}},\ \bibinfo {pages} {13} (\bibinfo {year}
  {1991})}\BibitemShut {NoStop}%
\bibitem [{\citenamefont {Angell}(1995)}]{angell1995}%
  \BibitemOpen
  \bibfield  {author} {\bibinfo {author} {\bibfnamefont {C.~A.}\ \bibnamefont
  {Angell}},\ }\href {\doibase 10.1126/science.267.5206.1924} {\bibfield
  {journal} {\bibinfo  {journal} {Science}\ }\textbf {\bibinfo {volume}
  {267}},\ \bibinfo {pages} {1924} (\bibinfo {year} {1995})}\BibitemShut
  {NoStop}%
\bibitem [{\citenamefont {Thiery}(2002)}]{thiery2002}%
  \BibitemOpen
  \bibfield  {author} {\bibinfo {author} {\bibfnamefont {J.~P.}\ \bibnamefont
  {Thiery}},\ }\href {\doibase 10.1038/nrc822} {\bibfield  {journal} {\bibinfo
  {journal} {Nat. Rev. Cancer}\ }\textbf {\bibinfo {volume} {2}},\ \bibinfo
  {pages} {442} (\bibinfo {year} {2002})}\BibitemShut {NoStop}%
\bibitem [{\citenamefont {Thiery}\ and\ \citenamefont
  {Sleeman}(2006)}]{thiery2006}%
  \BibitemOpen
  \bibfield  {author} {\bibinfo {author} {\bibfnamefont {J.~P.}\ \bibnamefont
  {Thiery}}\ and\ \bibinfo {author} {\bibfnamefont {J.~P.}\ \bibnamefont
  {Sleeman}},\ }\href {\doibase 10.1038/nrm1835} {\bibfield  {journal}
  {\bibinfo  {journal} {Nat. Rev. Mol. Cell Biol.}\ }\textbf {\bibinfo {volume}
  {7}},\ \bibinfo {pages} {131} (\bibinfo {year} {2006})}\BibitemShut {NoStop}%
\bibitem [{\citenamefont {Jolly}\ \emph {et~al.}(2015)\citenamefont {Jolly},
  \citenamefont {Boareto}, \citenamefont {Huang}, \citenamefont {Jia},
  \citenamefont {Lu} \emph {et~al.}}]{mohit2015}%
  \BibitemOpen
  \bibfield  {author} {\bibinfo {author} {\bibfnamefont {M.~K.}\ \bibnamefont
  {Jolly}}, \bibinfo {author} {\bibfnamefont {M.}~\bibnamefont {Boareto}},
  \bibinfo {author} {\bibfnamefont {B.}~\bibnamefont {Huang}}, \bibinfo
  {author} {\bibfnamefont {D.}~\bibnamefont {Jia}}, \bibinfo {author}
  {\bibfnamefont {M.}~\bibnamefont {Lu}},  \emph {et~al.},\ }\href {\doibase
  10.3389/fonc.2015.00155} {\bibfield  {journal} {\bibinfo  {journal} {Front.
  Oncol.}\ }\textbf {\bibinfo {volume} {5}},\ \bibinfo {pages} {155} (\bibinfo
  {year} {2015})}\BibitemShut {NoStop}%
\bibitem [{\citenamefont {Prost}\ \emph {et~al.}(2015)\citenamefont {Prost},
  \citenamefont {J{\"{u}}licher},\ and\ \citenamefont {Joanny}}]{jacques2015}%
  \BibitemOpen
  \bibfield  {author} {\bibinfo {author} {\bibfnamefont {J.}~\bibnamefont
  {Prost}}, \bibinfo {author} {\bibfnamefont {F.}~\bibnamefont
  {J{\"{u}}licher}}, \ and\ \bibinfo {author} {\bibfnamefont {J.~F.}\
  \bibnamefont {Joanny}},\ }\href {\doibase 10.1038/nphys3224} {\bibfield
  {journal} {\bibinfo  {journal} {Nat. Phys.}\ }\textbf {\bibinfo {volume}
  {11}},\ \bibinfo {pages} {111} (\bibinfo {year} {2015})}\BibitemShut
  {NoStop}%
\bibitem [{\citenamefont {Caprini}\ \emph
  {et~al.}(2020{\natexlab{a}})\citenamefont {Caprini}, \citenamefont {Marini
  Bettolo~Marconi},\ and\ \citenamefont {Puglisi}}]{caprini2020}%
  \BibitemOpen
  \bibfield  {author} {\bibinfo {author} {\bibfnamefont {L.}~\bibnamefont
  {Caprini}}, \bibinfo {author} {\bibfnamefont {U.}~\bibnamefont {Marini
  Bettolo~Marconi}}, \ and\ \bibinfo {author} {\bibfnamefont {A.}~\bibnamefont
  {Puglisi}},\ }\href {\doibase 10.1103/PhysRevLett.124.078001} {\bibfield
  {journal} {\bibinfo  {journal} {Phys. Rev. Lett.}\ }\textbf {\bibinfo
  {volume} {124}},\ \bibinfo {pages} {078001} (\bibinfo {year}
  {2020}{\natexlab{a}})}\BibitemShut {NoStop}%
\bibitem [{\citenamefont {Caprini}\ \emph
  {et~al.}(2020{\natexlab{b}})\citenamefont {Caprini}, \citenamefont {Marconi},
  \citenamefont {Maggi}, \citenamefont {Paoluzzi},\ and\ \citenamefont
  {Puglisi}}]{caprini2020b}%
  \BibitemOpen
  \bibfield  {author} {\bibinfo {author} {\bibfnamefont {L.}~\bibnamefont
  {Caprini}}, \bibinfo {author} {\bibfnamefont {U.~M.~B.}\ \bibnamefont
  {Marconi}}, \bibinfo {author} {\bibfnamefont {C.}~\bibnamefont {Maggi}},
  \bibinfo {author} {\bibfnamefont {M.}~\bibnamefont {Paoluzzi}}, \ and\
  \bibinfo {author} {\bibfnamefont {A.}~\bibnamefont {Puglisi}},\ }\href
  {\doibase 10.1103/PhysRevResearch.2.023321} {\bibfield  {journal} {\bibinfo
  {journal} {Phys. Rev. Res.}\ }\textbf {\bibinfo {volume} {2}},\ \bibinfo
  {pages} {023321} (\bibinfo {year} {2020}{\natexlab{b}})}\BibitemShut
  {NoStop}%
\bibitem [{\citenamefont {Chat\'e}\ \emph {et~al.}(2006)\citenamefont
  {Chat\'e}, \citenamefont {Ginelli},\ and\ \citenamefont
  {Montagne}}]{chate2006}%
  \BibitemOpen
  \bibfield  {author} {\bibinfo {author} {\bibfnamefont {H.}~\bibnamefont
  {Chat\'e}}, \bibinfo {author} {\bibfnamefont {F.}~\bibnamefont {Ginelli}}, \
  and\ \bibinfo {author} {\bibfnamefont {R.}~\bibnamefont {Montagne}},\ }\href
  {\doibase 10.1103/PhysRevLett.96.180602} {\bibfield  {journal} {\bibinfo
  {journal} {Phys. Rev. Lett.}\ }\textbf {\bibinfo {volume} {96}},\ \bibinfo
  {pages} {180602} (\bibinfo {year} {2006})}\BibitemShut {NoStop}%
\end{thebibliography}%

\end{document}